\documentclass[fleqn,usenatbib,useAMS]{mnras}

\usepackage{graphicx}	
\usepackage{amsmath}	
\usepackage{amssymb}	
\usepackage{multicol}        
\usepackage{bm}		
\usepackage{pdflscape}	
\usepackage{caption}

\newcommand{\angstrom}{\textup{\AA}}

\usepackage[T1]{fontenc}
\usepackage{ae,aecompl}

\usepackage{newtxtext,newtxmath}

\title[New Methods for Identifying LyC Leakers]{New Methods for Identifying Lyman Continuum Leakers and Reionization-Epoch Analogues }

\author[H. Katz]{Harley Katz$^{1}$\thanks{Visitor, Contact e-mail: \href{mailto:harley.katz@physics.ox.ac.uk}{harley.katz@physics.ox.ac.uk},}, Dominika \v{D}urov\v{c}\'{i}kov\'{a}$^{2}$, Taysun Kimm$^3$, Joki Rosdahl$^4$, \newauthor Jeremy Blaizot$^4$, Martin G. Haehnelt$^5$, Julien Devriendt$^1$, Adrianne Slyz$^1$, \newauthor Richard Ellis$^6$, \& Nicolas Laporte$^{5,7}$
\\
$^1$Astrophysics, University of Oxford, Denys Wilkinson Building, Keble Road, Oxford OX1 3RH, UK \\
$^2$New College, University of Oxford, Holywell Street, Oxford OX1 3BN, UK \\
$^3$Department of Astronomy, Yonsei University, 50 Yonsei-ro, Seodaemun-gu, Seoul 03722, Republic of Korea \\  
$^4$Univ Lyon, Univ Lyon1, Ens de Lyon, CNRS, Centre de Recherche Astrophysique de Lyon UMR5574, F-69230, Saint-Genis-Laval, France \\
$^5$Kavli Institute for Cosmology and Institute of Astronomy, Madingley Road, Cambridge CB3 0HA, UK \\
$^6$Department of Physics \& Astronomy, University College London, London, WC1E 6BT, UK \\
$^7$Cavendish Laboratory, University of Cambridge, 19 JJ Thomson Avenue, Cambridge CB3 0HE, UK}

\pubyear{2020}

\begin{document}
\label{firstpage}
\pagerange{\pageref{firstpage}--\pageref{lastpage}}
\maketitle

\begin{abstract}
Identifying low-redshift galaxies that emit Lyman Continuum radiation (LyC leakers) is one of the primary, indirect methods of studying galaxy formation in the epoch of reionization. However, not only has it proved challenging to identify such systems, it also remains uncertain whether the low-redshift LyC leakers are truly "analogues" of the sources that reionized the Universe. Here, we use high-resolution cosmological radiation hydrodynamics simulations to examine whether simulated galaxies in the epoch of reionization share similar emission line properties to observed LyC leakers at $z\sim3$ and $z\sim0$. We find that the simulated galaxies with high LyC escape fractions ($f_{\rm esc}$) often exhibit high O32 and populate the same regions of the R23-O32 plane as $z\sim3$ LyC leakers. However, we show that viewing angle, metallicity, and ionisation parameter can all impact where a galaxy resides on the O32-$f_{\rm esc}$ plane. Based on emission line diagnostics and how they correlate with $f_{\rm esc}$, lower-metallicity LyC leakers at $z\sim3$ appear to be good analogues of reionization-era galaxies. In contrast, local [SII]-deficient galaxies do not overlap with the simulated high-redshift LyC leakers on the SII-BPT diagram; however, this diagnostic may still be useful for identifying leakers. We use our simulated galaxies to develop multiple new diagnostics to identify LyC leakers using IR and nebular emission lines. We show that our model using only [CII]$_{\rm 158\mu m}$ and [OIII]$_{\rm 88\mu m}$ can identify potential leakers from non-leakers from the local Dwarf Galaxy Survey. Finally, we apply this diagnostic to known high-redshift galaxies and find that MACS1149\_JD1 at $z=9.1$ is the most likely galaxy to be actively contributing to the reionization of the Universe.  
\end{abstract}

\begin{keywords}
galaxies: high-redshift, (cosmology:) dark ages, reionization, first stars, galaxies: ISM, (ISM:) HII regions, galaxies: star formation
\end{keywords}




\section{Introduction}
While it is well established that the Universe transitioned from a nearly neutral to an ionised state during its early evolution, much remains uncertain about this process of reionization.  Reionization likely began at $z\gtrsim30$ with the formation of the first Population~III stars \citep{Wise2019}.  However, the redshift at which reionization ended is currently debated, with estimates ranging from $z\sim6$ \citep{Fan2006} to $z\sim5.3$ \citep{Kulkarni2019}.  In addition to the history of reionization, the sources that provided the photons that ionised the IGM remain unknown.  Numerous possibilities exist including dwarf galaxies and mini-haloes \citep{Couchman1986,Bouwens2015}, massive galaxies \citep{Naidu2020}, active galactic nuclei \citep{Haiman1998,Madau2015}, accretion shocks \citep{Dopita2011}, globular clusters \citep{Ricotti2002,Katz2013,Katz2014}, stellar mass black holes \citep{Madau2004,Ricotti2004,Mirabel2011}, and dark matter annihilation and decay \citep{Mapelli2006}.

Ideally, one could directly observe these objects during the epoch of reionization and simply measure the amount of LyC radiation that escapes the source and ionises the IGM.  However, in practice, this is incredibly difficult, if not impossible, because of the Gunn-Peterson Effect \citep{Gunn1965}, which causes the absorption of nearly all photons blue-ward of Ly$\alpha$ due to a neutral IGM.  Hence our understanding of the escape of ionising photons from their source populations into the IGM during the epoch of reionization comes from predominantly two methods: (1) Using numerical radiation hydrodynamics simulations to simulate the process from first principles \citep{Wise2014,Kimm2017,Rosdahl2018,Trebitsch2017} and (2) Using observations to study low-redshift "analogues" that are either leaking LyC photons or inferred to be leaking LyC radiation that may be representative of the high-redshift source population \citep[e.g.][]{Matthee2017,Borthakur2014,Leitherer2016,Vanzella2015,Siana2015,Shapley2016,Marchi2017,Naidu2018}.

Despite substantial differences in modelling approaches, certain trends have emerged from high-resolution numerical simulations regarding how LyC radiation escapes galaxies.  The general consensus is that this process is "feedback-regulated", indicating that strong supernova (SN), or perhaps AGN feedback, is required to clear channels from the centres of galaxies through which LyC radiation can escape \citep{Kimm2017,Trebitsch2017,Trebitsch2020}.  Galaxies are expected to evolve through "bursty" phases whereby the galaxy is most often opaque to LyC radiation and this only changes after a substantial star formation episode.  Simulations also tend to show that observing the escaping radiation is viewing angle-dependent \citep{Cen2015}.  Furthermore, galaxies of different masses do not leak LyC radiation equally and simulations tend to agree that low mass galaxies are much more efficient in this regard \citep{Wise2014,Xu2016,Paardekooper2015,Kimm2017} although see \citep[e.g.][]{Ma2020}. 

Observationally, identifying low-redshift systems with leaking LyC radiation has proved a difficult exercise, with most estimates of the LyC escape fraction ($f_{\rm esc}$), being less than what is required to reionize the Universe \citep{Siana2007,Siana2010,Bridge2010,Leitet2011,Leitet2013,Rutkowski2016,Leitherer2016}.  Knowing where to look is one of the key challenges and there has been considerable success at $z\gtrsim3$ \citep{Steidel2018,Nakajima2019,Fletcher2019,Mestric2020}. Low-redshift studies often focus on low-metallicity star-forming galaxies that have similar properties to higher-redshift Ly$\alpha$ emitters \citep[e.g.][]{Nakajima2016,Trainor2015,Dressler2015}.  Many of the leakers that have been observed can be characterised by certain emission line properties such as high line ratio of O32 ($\log_{10}({\rm [OIII]\ 4960\angstrom,\ 5007\angstrom/[OII]\ 3727\angstrom,\ 3729\angstrom})$) \citep{Nakajima2019,Faisst2016,Izotov2018a,debarros2016}, a small separation between the red and blue Ly$\alpha$ peaks \citep{Izotov2018a,Verhamme2017}, or deficits in SII emission \citep{Wang2019}.  Many of these trends can be understood using idealised models for HII regions \citep{Nakajima2014,Jaskot2013,Guseva2004} and by studying the differences between density-bounded and ionisation-bounded HII regions.  Nevertheless, it still remains to be determined whether the systems with observed LyC leakage at $z\lesssim6$ are actually representative of the high-redshift galaxies that may be reionizing the Universe.

In this work, we compare the emission line properties of low-redshift LyC leakers with those of simulated galaxies that are forming deep into the epoch of reionization to determine whether low-redshift "analogues" are representative of the high-redshift galaxy population. Furthermore, we use the emission line properties of our simulated galaxies to develop new methods for identifying LyC leaker "analogues" at low-redshift. Throughout this work, we define LyC leakers from non-leakers as galaxies that have an $f_{\rm esc}\geq10\%$. This separation is motivated by the fact that in order to complete reionization by $z\sim6$, one needs to adopt an escape fraction for galaxies in the range of $\sim7\%-20\%$ \citep{Robertson2015,Rosdahl2018}.

This paper is organised as follows. In Section~\ref{num_methods}, we introduce the Aspen simulations and how we model the LyC escape fraction and emission line luminosities. In Sections~\ref{results} and \ref{diagnostic}, we compare our simulated systems with observations and derive new diagnostics for identifying LyC leakers from observations. We test these diagnostics against observations at $z=0$ and $z>6$ to identify potential local LyC leakers as well as galaxies that played an active role in reionizing the Universe. Finally, in Sections~\ref{discussion} and \ref{conclusion}, we present our discussion and conclusions.

\section{Numerical Methods}
In this section, we describe our cosmological radiation hydrodynamics simulations as well as how we measure the escape fractions and emission lines from each of the simulated galaxies.

\label{num_methods}
\subsection{Cosmological Radiation Hydrodynamics Simulations}
For this work, we employ the state-of-the-art Aspen Suite of simulations \citep{Katz2019a,Katz2019b}, a set of cosmological radiation hydrodynamics simulations that zoom-in on the formation of a massive galaxy in the epoch of reionization.  Complete details of these simulations are presented in \cite{Katz2019b} and here, we briefly summarise the main properties of the simulations.

Our simulation zooms in on the formation of a Lyman-Break galaxy that has a dark matter halo mass of $10^{11.8}{\rm M_{\odot}}$ by $z=6$.  Initial conditions were generated for both dark matter and gas with MUSIC \citep{Hahn2011} at $z=150$ using a convex-hull shape for the high-resolution region.  We employ a \cite{Planck2016} cosmology with the following cosmological parameters: $h=0.6731$, $\Omega_{\rm m}=0.315$, $\Omega_{\rm b}=0.049$, $\Omega_{\rm \Lambda}=0.685$, $\sigma_8=0.829$, and $n_s=0.9655$.  The effective resolution of our simulation is $4,096^3$, corresponding to a dark matter particle mass of $4\times10^4{\rm M_{\odot}}h^{-1}$.  We use the Amiga Halo Finder \citep[AHF][]{Gill2004,Knollmann2009} to extract haloes from our simulation and we define the virial radius to be that which contains a mean density of dark matter, gas, and stars equal to $\Delta\rho_{\rm crit}$, where $\rho_{\rm crit}$ is the critical density of the Universe and $\Delta$ is the over-density that would allow for spherical collapse in an expanding cosmological background at a given redshift.  We consider only galaxies deep into the epoch of reionization where $\Delta\sim200$.  Furthermore, we only study haloes that are completely uncontaminated by low resolution dark matter particles and that are sampled by a minimum of 300 dark matter particles.  This corresponds to a minimum dark matter halo mass in our simulation of $1.2\times10^7{\rm M_{\odot}}h^{-1}$, which is well below the atom cooling threshold mass. By $z=10$, there are more than 1,000 uncontaminated haloes of at least this mass.  We assume an initial gas composition in the simulation of 76\%~H and 24\%~He by mass. 

Gravity, hydrodynamics, radiative transfer, and non-equilibrium chemistry are modelled using the version of RAMSES \citep{Teyssier2002,Rosdahl2013,Rosdahl2015} presented in \cite{Katz2017,Kimm2017}.  For the hydrodynamics, we use the MUSCL-Hancock scheme with an HLLC Riemann solver \citep{Toro1994} and a MinMod slope limiter \citep{Roe1986}.  The relation between gas pressure and internal energy is closed by assuming the gas is ideal and monatomic with $\gamma=5/3$.  Radiative transfer is solved with a first-order moment method (using the M1 closure \citep{Levermore1984} and a GLF intercell flux function \citep{Toro2009}).  We parametrise the radiation field with eight frequency bins from infrared (IR) to HeII-ionising, that are coupled to the gas thermochemistry via photoionisation, photo-heating, and radiation pressure (both direct UV and multi-scattered IR).  Non-equilibrium chemistry is solved for seven species: HI, HeI, HII, HeII, HeIII, H$_2$, and $e^-$.  To make the radiative transfer computational tractable, we use the reduced speed of light approximation and set the speed of light in the simulation to be $0.01c$.  Furthermore we sub-cycle the radiative transfer calculation up to 500 times per hydrodynamic time step \citep{Commercon2014,Rosdahl2013}.  We model the cooling of the gas in a non-equilibrium manner for the species listed above taking into account collisional ionisations, recombinations, collisional excitation, bremsstrahlung, Compton cooling and heating, and dielectronic recombination (see Appendix E of \citealt{Rosdahl2013}).  H$_2$ cooling is included using the rates from \cite{Hollenbach1979} as are equilibrium rates for metal line cooling \citep{Rosen1995}.

\begin{figure*}
\centerline{\includegraphics[scale=1,trim={0 0.0cm 0 0.0cm},clip]{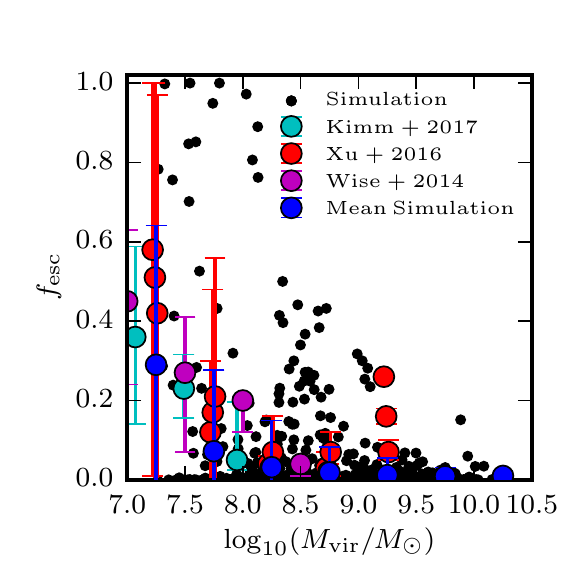}\includegraphics[scale=1,trim={0 0.0cm 0 0.0cm},clip]{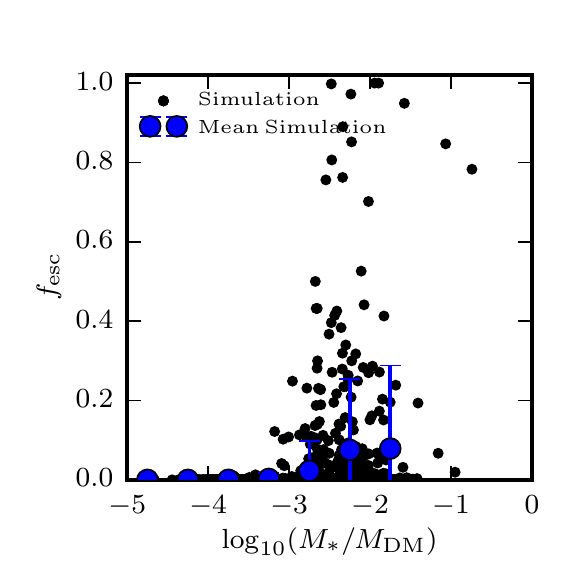}\includegraphics[scale=1,trim={0 0.0cm 0 0.0cm},clip]{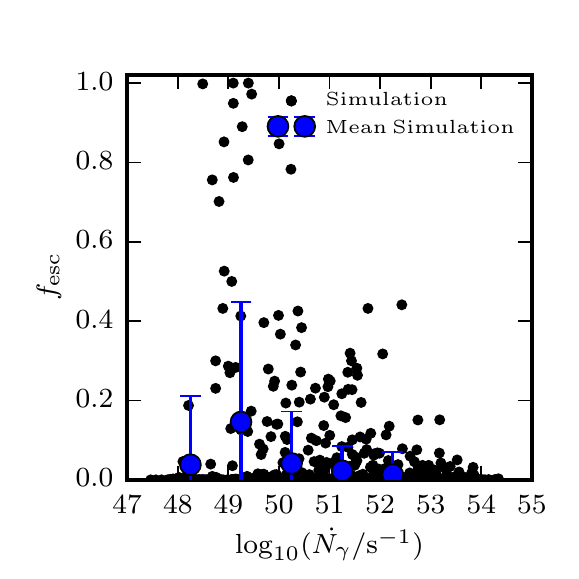}}
\caption{(Left) Virial mass of the galaxy versus LyC escape fraction at $12\geq z\geq9.2$. There are 9 simulation snapshots stacked in each of these plots.  Individual galaxies in our simulation are shown as black points while blue points with error bars represent mean values of $f_{\rm esc}$ within bins of 0.5~dex in virial mass.  For comparison we show the results from simulations of \protect\cite{Kimm2017} in cyan, \protect\cite{Xu2016} in red, and \protect\cite{Wise2014} in purple.  Despite differences in star formation routines, feedback schemes, and spatial and mass resolution, good agreement is seen among the different simulations. (Centre) Ratio of stellar mass to halo mass of the galaxy versus LyC escape fraction for individual galaxies in the simulation. (Right)  Instantaneous, intrinsic ionising luminosity versus LyC escape fraction for individual galaxies in the simulation.}
\label{global_props}
\end{figure*}

Star formation is modelled in the simulation based on a thermo-turbulent recipe \citep{Kimm2017,Rosdahl2018} and a Schmidt law \citep{Schmidt1959}.  The stellar mass in our simulations is set to a minimum of $1,000{\rm M_{\odot}}$.  Star particles inject radiation into their host cell based on their mass, age, and metallicity, according to a BPASSv2.0 SED \citep{Eldridge2008,Stanway2016} assuming an IMF with a maximum stellar mass of 300${\rm M_{\odot}}$ and a slope of $-1.30$ between 0.1 and 0.5${\rm M_{\odot}}$ and $-2.35$ between 0.5 and 300${\rm M_{\odot}}$.  For the first 50Myr during the lifetime of each star particle, they can undergo SN which are randomly drawn from a realistic delay-time distribution \citep{Kimm2015}.  SN are modelled using the mechanical scheme of \cite{Kimm2015} and the equivalent of $10^{51}$ergs is injected back into the gas for each SN.  In the case where the HII region around a star particle is unresolved, a fixed value of $5\times10^5~{\rm M_{\odot}km/s}$ of momentum per SN is injected \citep{Kimm2017,Geen2015}.  For each SN, 20\% of the mass of the star is recycled back into the gas assuming it is metal enriched to a value of 0.075 \citep{Kroupa2001}.  We employ a 4$\times$ boost to the number of SN, consistent with \cite{Rosdahl2018} to match the high-redshift extrapolation of the stellar mass-halo mass relation.  This also produces a reasonable reionization history and UV luminosity function \citep{Rosdahl2018}.

Because we use an AMR technique for our simulation, we use a quasi-Lagrangian technique to refine cells.  When a cell contains either eight dark matter particles or the gas mass of the cell is $>8\frac{\Omega_{\rm b}}{\Omega_{\rm DM}}{\rm m_{DM}}$, where ${\rm m_{DM}}$ is the dark matter particle mass, the cell is refined.  Similarly, a cell is also refined if the cell width is more than $1/4$ of the local Jeans length.  We allow the simulation to refine to maintain a constant physical resolution of 13.6pc.

For full details of the simulation, please refer to \cite{Katz2019a,Katz2019b}.

\subsection{Nebular and IR Emission Line Luminosities}
In order to compare with observations, we need to measure different emission line luminosities for each galaxy in our simulation.  We do this in post-processing by employing the photoionisation code CLOUDY \citep{Ferland2017}.  For each cell in the simulation, we know the temperature, density, metallicity, and local radiation field (strength and shape).  Ideally, we would run a CLOUDY model for each cell; however, this is currently computationally unfeasible due to the total number of cells in the simulation and the number of snapshots that we wish to post-process.  We estimate that this would require at least a few million CPU hours.  For this reason, we employ a Random Forest \citep{Ho1995,Breiman2001,Geurts2006}, which is an ensemble machine learning technique to quickly estimate the various emission line luminosities that we are interested in.  The Random Forest was trained on 850,000 cells that were extracted from the $z=10$ snapshot of the simulation, and in \cite{Katz2019b}, we demonstrated that by using our trained model, we can estimate the total luminosities of galaxies in our simulation to better than 5\%-10\% accuracy for most luminous lines\footnote{This accuracy is already better than the systematic uncertainty of changing the geometry of the calculation (e.g. switching from a slab to a sphere which can change the results on the order of a few percent).} at a fraction of the computational cost of running CLOUDY.  Full details of the CLOUDY models and their accuracies are presented in \cite{Katz2019b}. For this work, we will only study the intrinsic line luminosities for each galaxy (not including dust attenuation), measured by summing the luminosity of each gas cell within the virial radius of each halo.  Furthermore, we only consider IR and nebular emission lines and will not study resonant lines (such as Ly$\alpha$) although they are indeed very interesting in the context of estimating the escape fraction \citep{Verhamme2015,Verhamme2017,Dijkstra2016,Kimm2019}.

\subsection{Measuring the Escape Fraction}
To measure the escape fraction ($f_{\rm esc}$) of galaxies in our simulations we post-process the simulations with a ray-tracing method.  Methods such as these have been shown to be equally good as directly measuring the ionising flux emanating from the galaxies \citep{Trebitsch2017}. For each galaxy in the simulation, we extract a sphere around the centre of each halo with radius equal to the virial radius of the system, and we project the neutral hydrogen density in each cell onto a cube (with a side length equal to the diameter of the halo) that has a cell width equal to the maximum physical resolution of the simulation in that snapshot ($\sim$13.6pc).  For each star particle in the halo, we then ray-trace \citep{Buchner2017} through the cube in 500 random directions from the location of each star particle, keeping the ionisation state of the gas fixed, to measure the optical depth to neutral hydrogen and thus the escape fraction. We only consider neutral hydrogen here and we assume a mean photon energy of $21$eV, which is the mean energy of a photon across the age and metallicity dependent SEDs that we employ, up to a metallicity of 0.5$Z_{\odot}$.  We then take a luminosity weighted average of each star particle to measure the total $f_{\rm esc}$ of each galaxy.  Filling the grid with dust, molecular hydrogen, and helium and ray tracing at multiple photon energies does not substantially change our measured $f_{\rm esc}$ and our method is significantly computationally cheaper (see Figure~5 of \citealt{Kimm2019}).  We have also checked that 500 rays is sufficient by measuring the differences between using between 100 and 100,000 rays.  

We measure $f_{\rm esc}$ for all galaxies in our simulation that host at least 20 star particles (i.e. that have a minimum stellar mass of $2\times10^4{\rm M_{\odot}}$).  Furthermore, in this work, we will only consider the global $f_{\rm esc}$ (i.e. that averaged over all directions) so that it can be compared to the intrinsic line luminosities.  It should be noted that in observations, $f_{\rm esc}$ can change quite substantially depending on viewing angle \citep{Cen2015}.  However, our method is more useful for determining how the physics of leaking radiation is related to emission line luminosities, without the complications of inhomogeneous dust and neutral hydrogen distributions. We note that if the distribution of dust is significantly different from that of the neutral hydrogen in real galaxies, observations will be impacted in a different way from the simulations where the dust is defined to be in the metal enriched, mostly neutral gas. 

\section{Results}
\label{results}
In this Section, we discuss how the results from our simulations compare to others in the literature and then use our simulations to deduce whether observed galaxies that have either high values of O32 or [SII] deficits are analogues of reionization-era galaxies.

\subsection{Global Properties}

In this work, we aim to elucidate whether the properties of low-redshift LyC leakers are consistent with those during the epoch of reionization as well as derive diagnostics for identifying the LyC leakers at low-redshift. Using the Aspen simulations, we have analysed the properties of all galaxies at nine different redshifts, deep into the epoch of reionization, in the range $12\geq z\geq 9.2$. In Figure~\ref{global_props}, we plot $f_{\rm esc}$ versus the virial mass of the halo for all galaxies in our simulation at all of the sampled redshifts. Consistent with previous work \citep{Wise2014,Xu2016,Kimm2017}, we identify a strong trend of $f_{\rm esc}$ with mass such that the lower mass systems exhibit higher average escape fractions compared to higher mass objects.  The individual black points represent individual galaxies in our simulation while the blue points represent the mean values in bins of virial mass with a width of 0.5~dex.  Despite differences in feedback models, resolution, and redshifts analysed, our escape fractions are consistent with the trend in other work in the literature \citep{Wise2014,Xu2016,Kimm2017}.

Simulations now tend to agree that the escape of LyC photons is feedback-regulated and that a strong burst of star formation combined with effective SN feedback is required for such photons to escape \citep[e.g.][]{Kimm2014,Trebitsch2017,Kimm2017,Rosdahl2018,Ma2020}. One may expect that the galaxies that have a high ratio of stellar mass to dark matter mass are the ones that likely have strong enough feedback to disrupt the gas in the centre of the halo and create channels for LyC photons to escape.  In the centre panel of Figure~\ref{global_props}, we plot the ratio of $M_*/M_{\rm DM}$ versus $f_{\rm esc}$.  In general, only the haloes with higher $M_*/M_{\rm DM}$ exhibit significant LyC leakage; however, we note that this could be impacted by simulation resolution as the higher resolution simulations of \cite{Kimm2017} demonstrate that certain galaxies with low $M_*/M_{\rm DM}$ can exhibit significant LyC leakage. Lower spatial resolution can smooth the ISM, which may inhibit LyC photons from escaping. Not all systems with high $M_*/M_{\rm DM}$ exhibit high $f_{\rm esc}$ as the timing of when the stars formed also matters. Since $f_{\rm esc}$ fluctuates, the time period to catch the phase of high $f_{\rm esc}$ after a star formation event can be as short as $1-10$Myr (see e.g. \citealt{Trebitsch2017,Kimm2017,Yoo2020}) and thus not all systems with high $M_*/M_{\rm DM}$ can be expected to have high $f_{\rm esc}$ at the fixed redshifts that we analyse. This timing argument also applies to the lowest mass systems with low $M_*/M_{\rm DM}$.  These galaxies have the fewest number of star particles and thus the probability of catching them in an optically thin phase is low.

In the right panel of Figure~\ref{global_props}, we compare the instantaneous production rate of ionising photons ($\dot{N}_{\gamma}$) with $f_{\rm esc}$.  There is considerable scatter in this plot; however, high escape fractions are preferentially found in less luminous galaxies in LyC. This is reflective of the fact that the lower mass objects in our simulations tend to have higher $f_{\rm esc}$ as there is a strong trend between virial mass and $f_{\rm esc}$. Certain highly luminous systems also have high $f_{\rm esc}$, many of which represent lower mass galaxies that have undergone a recent burst of star formation. 

The global properties of the galaxies in our simulation seem to agree well with previous work \citep{Kimm2017,Wise2014,Xu2016}, both in the context of their escape fraction properties as well as their star formation and emission line properties \citep{Katz2019b}. In subsequent sections, we relate the observational properties of our simulated galaxies with those of low-redshift LyC leakers, with an emphasis on nebular and IR emission lines.

\begin{figure}
\centerline{\includegraphics[scale=1,trim={0 0.0cm 0 0.0cm},clip]{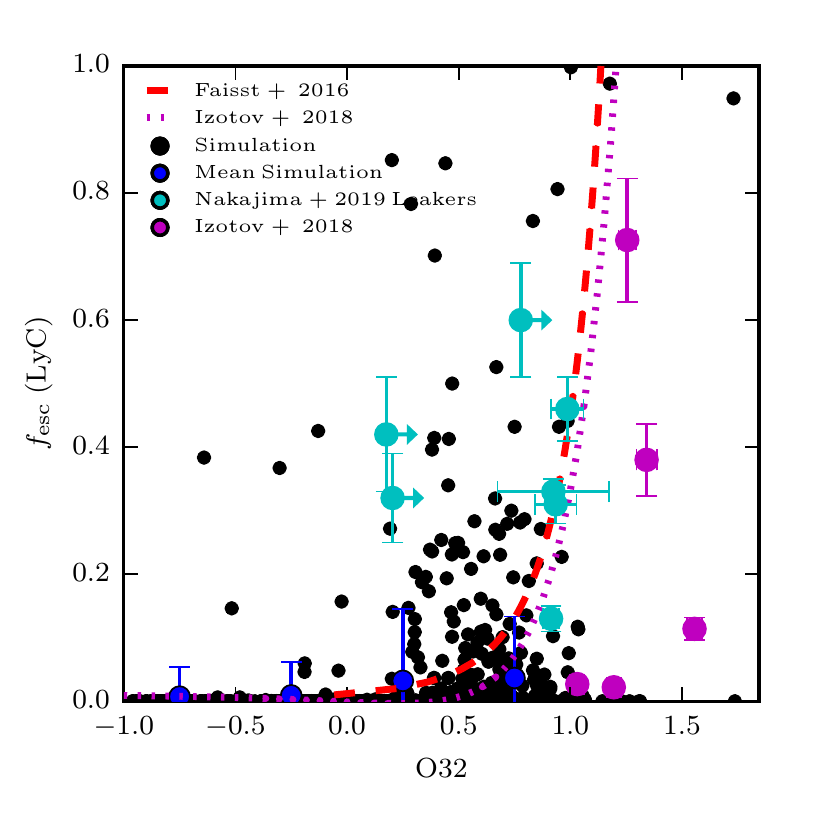}}
\caption{O32 ($\log_{10}({\rm [OIII]\ 4960\angstrom,\ 5007\angstrom/[OII]\ 3727\angstrom,\ 3729\angstrom})$) versus $f_{\rm esc}$ for all galaxies in our simulation (black points).  Cyan and purple points represent observed low-redshift LyC leakers from \protect\cite{Nakajima2019} and \protect\cite{Izotov2018a}, respectively. The blue dashed and purple dotted lines show the expected trends between O32 and $f_{\rm esc}$ from \protect\cite{Faisst2016} and \protect\cite{Izotov2018a}, respectively. In general, we find that systems that have a significant amount of LyC leakage also exhibit higher O32.}
\label{O32fesc}
\end{figure}

\subsection{O32}
One method that has been used for identifying LyC leakers at low redshift is to focus on systems that have excess [OIII] emission compared to [OII] emission (O32, $\log_{10}({\rm [OIII]\ 4960\angstrom,\ 5007\angstrom/[OII]\ 3727\angstrom,\ 3729\angstrom})$) \citep[e.g.][]{Nakajima2014}.  From a theoretical viewpoint, O32 correlates with ionisation parameter, the ratio of ionising photon density to hydrogen density \citep{Kewley2002}, although there is an additional dependency on metallicity. Galaxies with high ionisation parameters are more likely to be leaking ionising photons, hence O32 is expected to be a useful diagnostic for identifying LyC leakers.  Many observational works have confirmed that galaxies that exhibit leaking LyC radiation often have high O32 \citep{Nakajima2014,debarros2016,Izotov2018b,Faisst2016,Nakajima2019}. 

In Figure~\ref{O32fesc}, we plot O32 versus $f_{\rm esc}$ for our simulated galaxies compared with low-redshift observations. Consistent with the low-redshift observations, we find that the simulated galaxies that exhibit a significant fraction of LyC leakage also tend to be biased towards high O32. All of the observed galaxies that have high $f_{\rm esc}$ have ${\rm O32}>0$.  If we use ${\rm O32}=0$ as our dividing line between high and low O32, we find that in our sample of simulated galaxies, 11\% of systems with high O32 have $f_{\rm esc}>10\%$.  In contrast, only 1\% of the simulated galaxies with ${\rm O32}<0$ have $f_{\rm esc}>10\%$. The trend lines from \cite{Faisst2016} and \cite{Izotov2018a} represent reasonable approximations of our simulated galaxies although they tend to be slightly biased towards higher O32 for a given $f_{\rm esc}$. However, it is unclear if these trend lines match the simulations at the highest values of O32 where the simulated galaxies are sparsely sampled.  There is considerable scatter in this relation expected from both the observations and our simulations and it is also important to note that our work employs the angle averaged quantities for these values while observations can only probe a single line-of-sight. 

\begin{figure*}
\centerline{\includegraphics[scale=1,trim={0 0.0cm 0 0.0cm},clip]{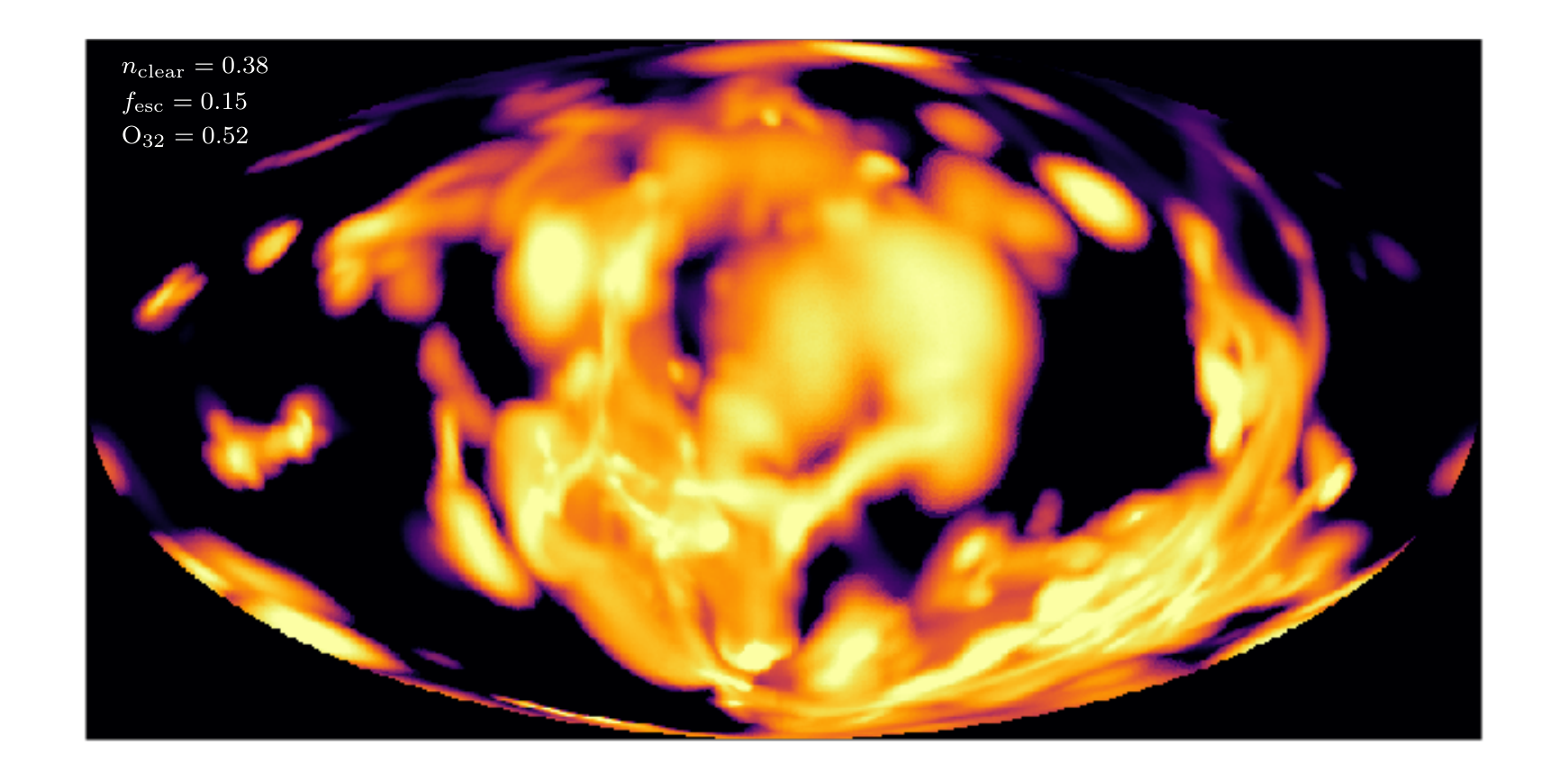}}
\centerline{\includegraphics[scale=1,trim={0 0.0cm 0 0.0cm},clip]{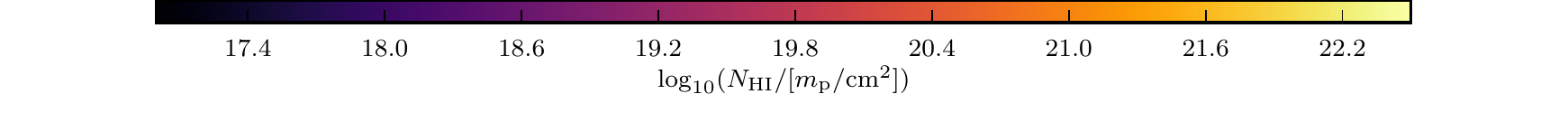}}
\centerline{\includegraphics[scale=1,trim={0 0.0cm 0 0.0cm},clip]{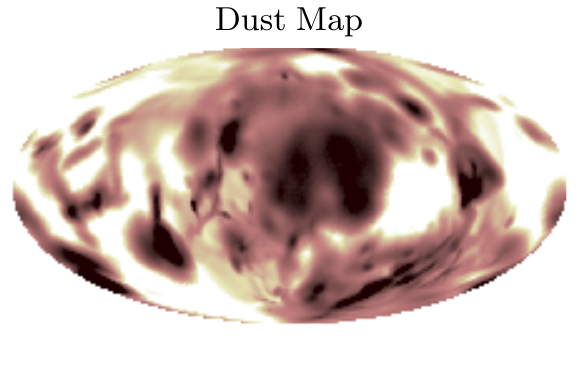}\includegraphics[scale=1,trim={0 0.0cm 0 0.0cm},clip]{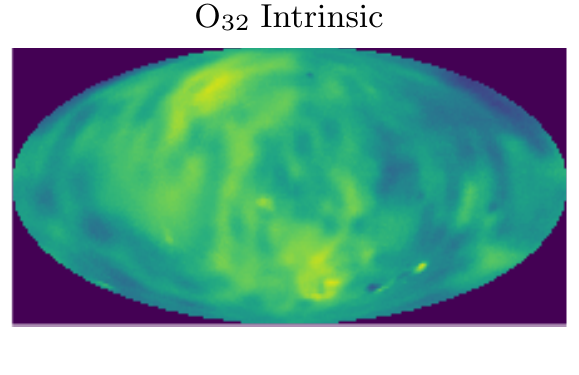}\includegraphics[scale=1,trim={0 0.0cm 0 0.0cm},clip]{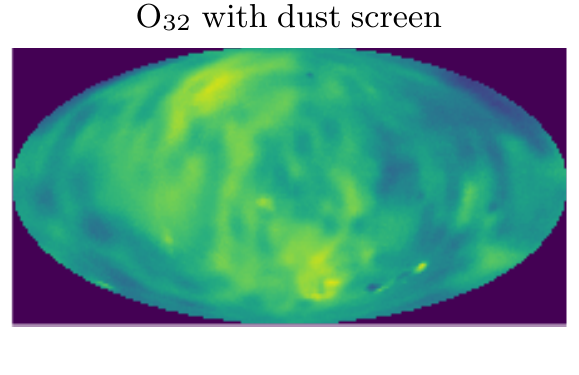}}
\centerline{\includegraphics[scale=1,trim={0 0.0cm 0 0.0cm},clip]{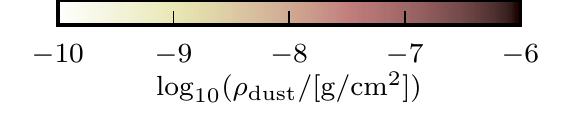}\includegraphics[scale=1,trim={0 0.0cm 0 0.0cm},clip]{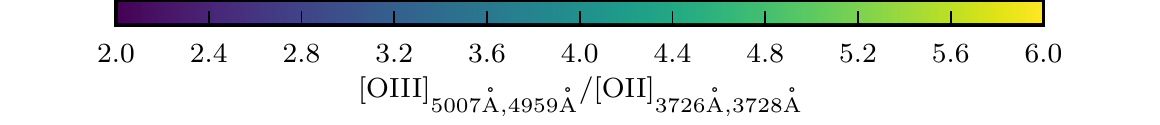}}
\caption{(Top) Molleweide projection of neutral hydrogen column density as observed from the centre of a galaxy with high O32 and leaking LyC radiation. This galaxy has a dark matter mass of $10^{9.9}\rm{M_{\odot}}$ and a stellar mass of $10^{7.9}\rm{M_{\odot}}$.  $n_{\rm clear}$ represents the fraction of sight lines that have an escape fraction of at least 10\%.  Depending on the viewing angle, an observer can either view an escape fraction of 0 or up to 100\%. (Bottom) Molleweide projections of the dust column density (left), intrinsic O32 (center), and O32 accounting for dust (right).  The dust tends to follow the locations of neutral hydrogen while O32 variations are more sensitive to the strength of the radiation field and feedback along a specific line-of-sight.  O32 is shown in linear scale on these images and different lines-of-sight show differences of a factor of three in O32. This galaxy is low metallicity and even though the dust distribution is highly inhomogeneous, the impact on O32 is only $\sim5\%$.}
\label{viewing_angle}
\end{figure*}

Recently, \cite{Bassett2019} have called into question the utility of this diagnostic as not all observed galaxies with high O32 also have high $f_{\rm esc}$. It is clear from Figure~\ref{O32fesc} that this is indeed the case in our simulations. \cite{Nakajima2019} have also identified a sample of observed galaxies at $z\sim3$ that have high O32 and low $f_{\rm esc}$ and attribute this to the impact of viewing angle. In contrast, \cite{Bassett2019} outline numerous idealised models of HII regions that show how the relation between O32 and $f_{\rm esc}$ responds to variations in metallicity and ionisation parameter. The results from our simulations demonstrate that LyC leakers are preferentially biased towards having high O32 and below, we study both effects outlined in the literature.

\begin{figure*}
High $f_{\rm esc}$, High ${\rm O32}$
\centerline{\includegraphics[scale=1,trim={0 0.0cm 0 0.0cm},clip]{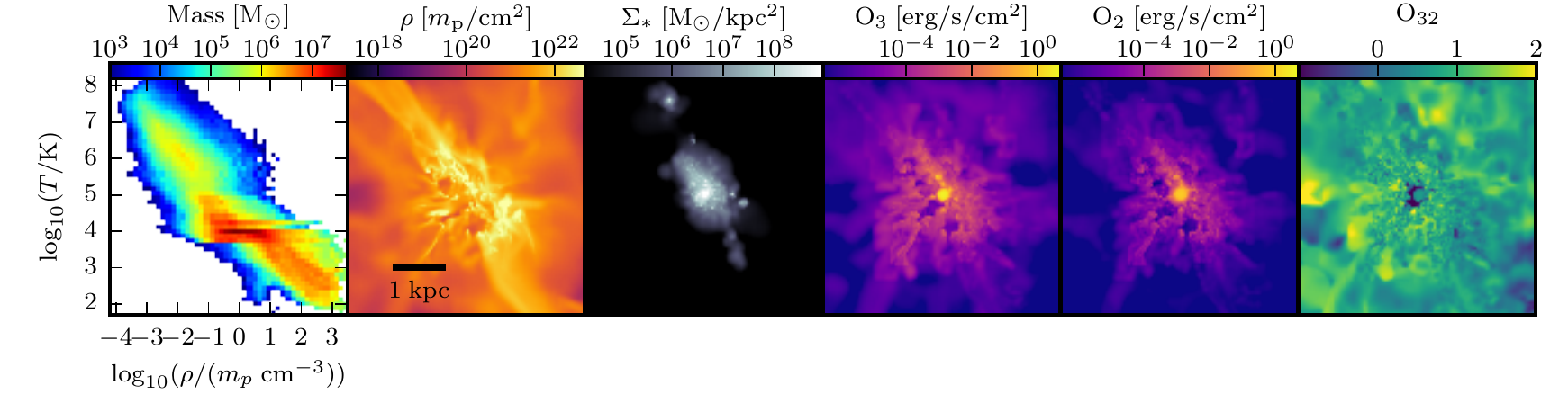}}
Low $f_{\rm esc}$, High ${\rm O32}$
\centerline{\includegraphics[scale=1,trim={0 0.0cm 0 0.0cm},clip]{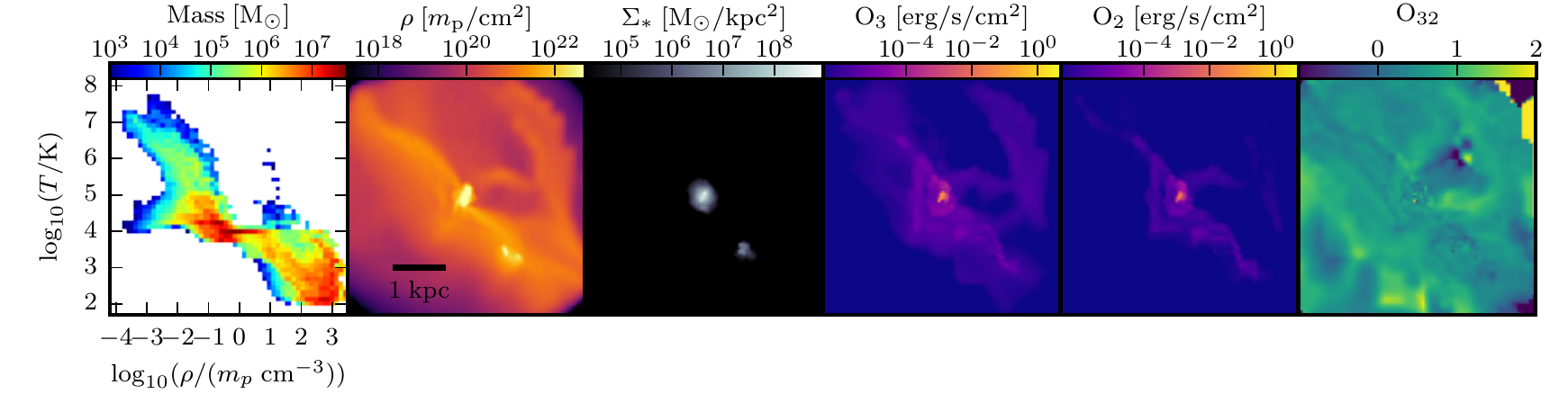}}
\caption{From left to right: Mass-weighted temperature-density phase-space diagram, gas column density map, stellar surface-mass density map, O3 intensity map, O2 intensity map, and spatial distribution of O32.  The top panel shows an example of a galaxy with high ${\rm O32}$ and high $f_{\rm esc}$ (Halo~1) while the bottom panel shows and example of a galaxy with high ${\rm O32}$ and low $f_{\rm esc}$ (Halo~2).  The properties of Halo~1 and Halo~2 can be found in Table~\ref{halo_props_table}. Individual regions of a galaxy might have very high or very low O32 depending on local conditions and deviate strongly from the value averaged across the galaxy. Despite both having high ${\rm O32}$, structural, metallicity, stellar population, and feedback differences between the two systems mean that they have very different LyC escape fractions.}
\label{metal_effect}
\end{figure*}

\begin{table*}
    \centering
    \begin{tabular}{lcccccc}
    \hline
    Halo & $f_{\rm esc}$ & $M_{\rm halo}$ & $M_{*}$ & SFR & Metallicity  & O32 \\
    \hline
        Halo 1 & 15.1\% & $7.66\times10^9{\rm M_{\odot}}$ & $7.64\times10^7{\rm M_{\odot}}$ & 0.24 ${\rm M_{\odot}yr^{-1}}$ & 0.030$Z_{\odot}$ & 0.52 \\
        Halo 2 & 0.0014\% & $1.76\times10^9{\rm M_{\odot}}$ & $6.57\times10^6{\rm M_{\odot}}$ & 0.08 ${\rm M_{\odot}yr^{-1}}$ & 0.017$Z_{\odot}$ & 0.67 \\
    \hline
    \end{tabular}
    \caption{Properties of the haloes shown in Figure~\ref{metal_effect}}
    \label{halo_props_table}
\end{table*}

\subsubsection{Impact of Viewing Angle}
\cite{Nakajima2019} recently studied a sample of Ly$\alpha$ emitters at $z\sim3.1$ and found that although the galaxies leaking LyC radiation were biased towards higher O32, this was not a sufficient condition for the galaxy to be leaking radiation.  Because there were no other identified spectral differences between the galaxies that were leakers versus non-leakers, they suggested that viewing angle was a possible explanation.  

In the top panel of Figure~\ref{viewing_angle}, we show a Molleweide projection of neutral hydrogen column density for a galaxy that is leaking LyC photons ($f_{\rm esc}=0.15$) and that has a higher O32 compared to many other systems in our sample (O32=0.52). This halo has a dark matter mass of $10^{9.9}\rm{M_{\odot}}$ and a stellar mass of $10^{7.9}\rm{M_{\odot}}$. We define $n_{\rm clear}$ to be the fraction of viewing angles where an observer would measure a $f_{\rm esc}>10\%$ and for this system, we find $n_{\rm clear}=38\%$. We would expect that most of the time, an observer would measure a high O32 but a low $f_{\rm esc}$. This figure demonstrates that for a given galaxy, the measured LyC $f_{\rm esc}$ is strongly dependent on viewing angle \citep[see also][]{Cen2015}.  Hence, for observed galaxies, viewing angle can significantly impact the trend between O32 and $f_{\rm esc}$, consistent with the arguments made in \cite{Nakajima2019}.

In addition to angular variations in $f_{\rm esc}$, it is plausible that there are also angular variations in O32. According to \cite{Bassett2019}, dust is not expected to significantly impact O32, with corrections expected to be $<0.3$ dex.  In the bottom left panel of Figure~\ref{viewing_angle}, we show a Molleweide projection of the dust column density.  The dust mass in each simulation cell is computed following the description in \cite{Katz2019a} where we apply a metallicity dependent dust-to-metal ratio \citep{RR2014} and assume that all dust has been destroyed in cells that have $T>10^5$K. The dust distribution is clearly very inhomogeneous and tends to follow the location of neutral hydrogen.  In the bottom center panel of Figure~\ref{viewing_angle}, we show line-of-sight variations in O32 (in linear scale) that differ by a factor of three.  In the bottom right panel, we show the dust attenuated O32 Molleweide projection (assuming that the inhomogeneous dust map acts as a dust screen for each line\footnote{Assuming that the dust distribution acts as a screen would represent the maximum impact of dust on O32.}) and we only find $\sim5\%$ differences accounting for dust.  Most of the leakers in our simulation have low metallicity so it is unlikely that for the current work, viewing angle will significantly impact O32.  In contrast, much higher metallicity galaxies could see a larger effect as indicated in \cite{Bassett2019}.

\subsubsection{Impact of Metallicity and Ionisation Parameter}
While viewing angle can certainly impact the observed relation between O32 and $f_{\rm esc}$, from Figure~\ref{O32fesc}, it is clear that there is a population of high-redshift galaxies that exhibit high ${\rm O32}$ but low $f_{\rm esc}$ that are not viewing angle dependent.  This is because, for this figure, we show the angle-averaged $f_{\rm esc}$, rather than that along a specific line-of-sight.  This opens the question of what intrinsic properties cause galaxies to have high ${\rm O32}$ but low $f_{\rm esc}$?

\begin{figure}
\centerline{\includegraphics[scale=1,trim={0 0.0cm 0 0.0cm},clip]{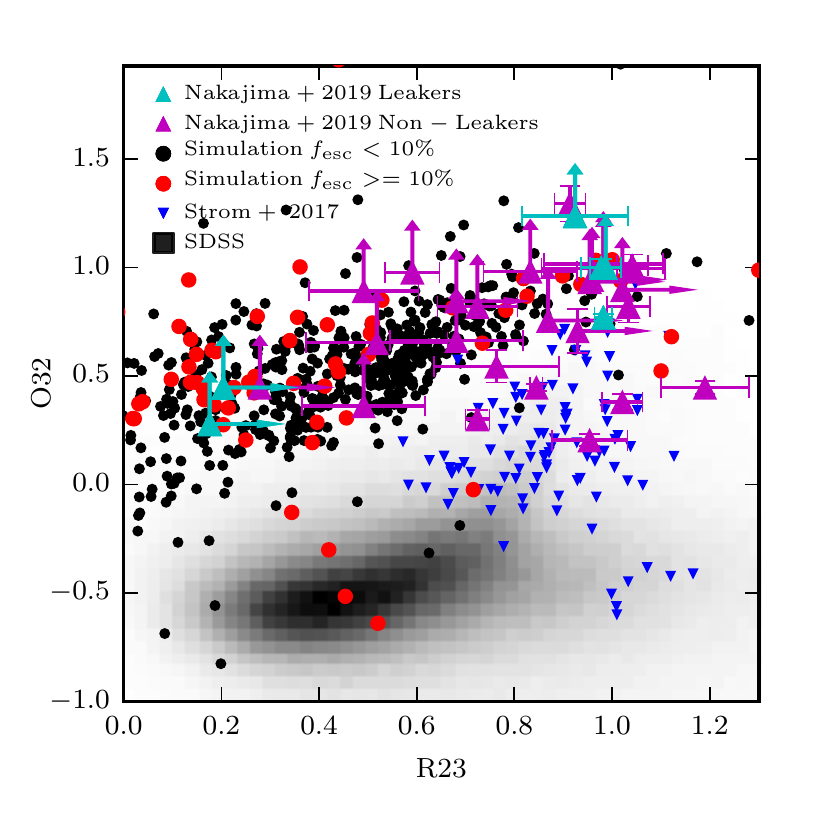}}
\caption{O32 versus R23 for galaxies in our simulation compared to observations.  The black and red points represent simulated galaxies with $f_{\rm esc}<10\%$ and $f_{\rm esc}\geq10\%$, respectively.  The shaded 2D histogram represents local galaxies as observed by SDSS.  Blue points represent $z\sim2-3$ galaxies from \protect\cite{Strom2017}, while the cyan points show the $z\sim3$ LyC leakers from \protect\cite{Nakajima2019}. We also show the non-leakers from  \protect\cite{Nakajima2019} in magenta.  Our simulated galaxies overlap similar regions on this diagram to the $z\sim3$ galaxies from \protect\cite{Nakajima2019}; however, this plane is not useful for classifying LyC leakers from non-leakers.}
\label{R23O32}
\end{figure}

In Figure~\ref{metal_effect} we show examples of two galaxies, one with high ${\rm O32}$ and high $f_{\rm esc}$ (top) and the other with high ${\rm O32}$ but low $f_{\rm esc}$ (bottom). The properties of these two galaxies are listed in Table~\ref{halo_props_table}. We have selected these two massive systems in haloes with mass $>10^9{\rm M_{\odot}}$ so that they are sufficiently resolved by our simulation.  For each galaxy, we show the mass-weighted temperature-density phase diagram, a map of gas column density, a map of stellar surface-mass density, maps of ${\rm [OIII]_{4960\angstrom+5007\angstrom}}$ and ${\rm [OII]_{3727\angstrom,3729\angstrom}}$ intensity, and a map of the O32 spatial distribution, from left to right.  There are clear and obvious structural differences between the two systems.  The system with high ${\rm O32}$ and high $f_{\rm esc}$ has both an extended gas and stellar distribution and there is clear evidence that stellar feedback and other dynamical effects have disturbed the structure of the galaxy.  This is evident in the phase diagram where significant amounts of ionised gas are present and likewise there is a considerable amount of gas within the virial radius that has a low density and a high temperature, a consequence of effective supernova feedback.  In contrast, the system with high ${\rm O32}$ and low $f_{\rm esc}$ is very compact, is not sufficiently impacted by stellar feedback, as seen from the phase diagram, and has very little ionised gas.  Much of this behaviour is driven by the fact that the first galaxy has a considerably higher stellar mass.  Although usually discussed in the context of HII regions, from a galaxy perspective, the first system with high ${\rm O32}$ and high $f_{\rm esc}$ can be thought of as a density bounded galaxy (i.e. when there is not enough gas to absorb all of the ionising radiation) while the second system with high ${\rm O32}$ and low $f_{\rm esc}$ can be thought of as an ionisation bounded galaxy (i.e. when the source of radiation fills out it's Stromgren sphere inside the cloud).  

The system with low $f_{\rm esc}$ has about half the metallicity and an order of magnitude lower stellar mass compared to the galaxy with high $f_{\rm esc}$.  From the photoionisation models of \cite{Bassett2019} (see the right panel of their Figure~10), it is clear that for fixed ionisation parameter and $f_{\rm esc}$, low-metallicity systems have systematically higher O32, likely caused by low-metallicity SEDs being harder.  This also causes our simulated systems to fall significantly higher on the R23-O32 (${\rm R23}=\log_{10}(([{\rm OIII}]\ 4960\angstrom,\ 5007\angstrom+[{\rm OII}]\ 3727\angstrom,\ 3729\angstrom)/{\rm H\beta})$) diagram compared to SDSS galaxies (see Figure~\ref{R23O32}), yet still consistent with observed systems at $z\sim3$ \citep{Nakajima2019,Katz2019b,Strom2017}. One can see from Figure~\ref{R23O32} that this plane is not particularly useful for differentiating between LyC leakers and non-leakers. Furthermore, in the models of \cite{Bassett2019}, for fixed ionisation parameter, the upturn in $f_{\rm esc}$ as a function of O32 occurs at systematically higher O32 as metallicity decreases.  The galaxies in our simulation have varying metallicities and in our two example galaxies, the trends with $f_{\rm esc}$ follow exactly that predicted by \cite{Bassett2019}.  It is important to emphasise that ionisation parameter (and also strength of stellar feedback) also matters in determining $f_{\rm esc}$.  

\begin{figure}
\centerline{\includegraphics[scale=1,trim={0 0.6cm 0 0.8cm},clip]{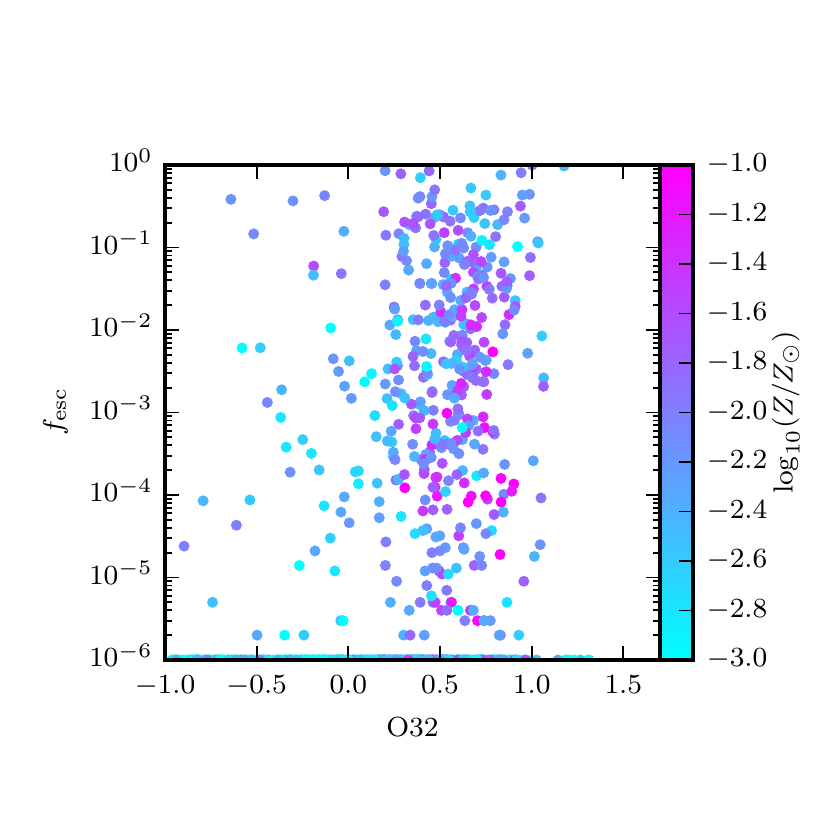}}
\centerline{\includegraphics[scale=1,trim={0 0.6cm 0 0.8cm},clip]{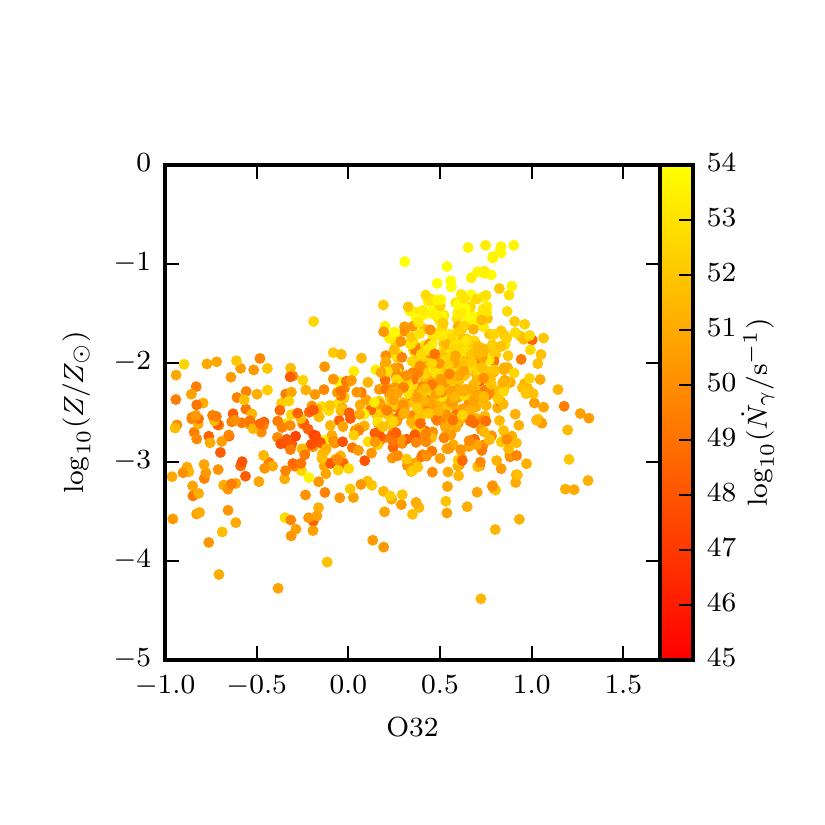}}
\centerline{\includegraphics[scale=1,trim={0 0.6cm 0 0.8cm},clip]{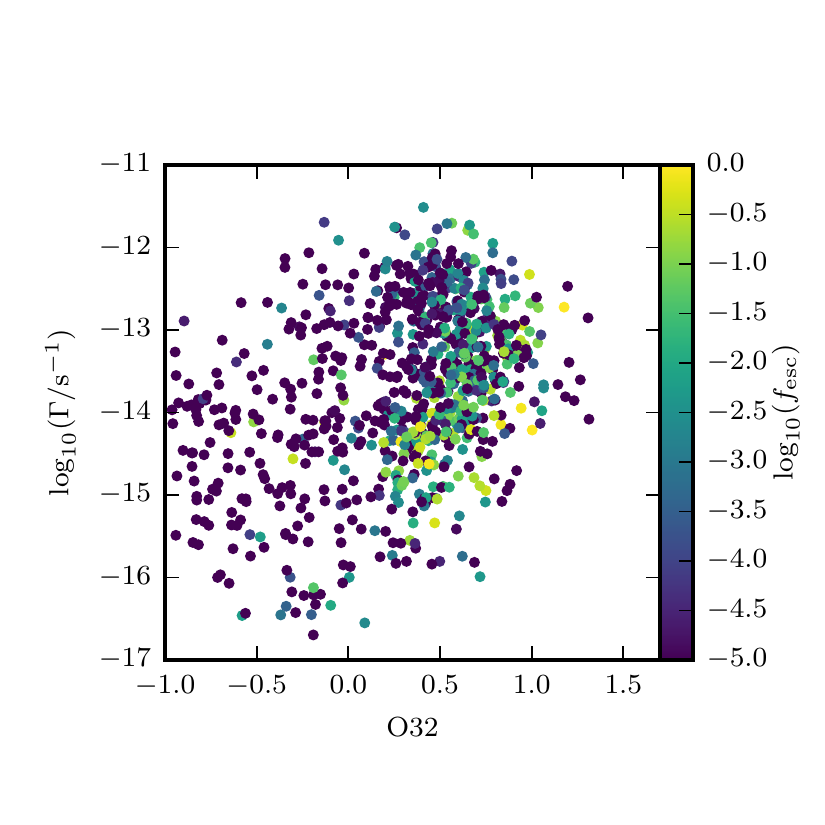}}
\caption{(Top) O32 versus $f_{\rm esc}$ for galaxies in our simulation coloured by metallicity. All galaxies with $f_{\rm esc}<10^{-6}$ have been set to that value so they are visible on the plot. (Middle) O32 versus metallicity for galaxies in our simulation coloured by ionising luminosity.  (Bottom) O32 versus HI ionisation rate in the halo for galaxies in our simulation coloured by escape fraction.  While it is difficult to identify trends in the O32-$f_{\rm esc}$ plane, it is clear that the galaxies in our simulation with the highest O32 (e.g. ${\rm O32}>1$) have lower metallicities and higher HI ionisation rates.}
\label{metal_effect2}
\end{figure}

In the top panel of Figure~\ref{metal_effect2} we show O32 versus $f_{\rm esc}$ where we have colour-coded the points by the metallicity of the galaxy. In the middle panel of Figure~\ref{metal_effect2}, we plot O32 versus galaxy metallicity colour-coded by ionising luminosity.  From the middle panel, we see that the systems at fixed O32 with higher metallicity also tend to have higher ionising luminosities.  This is a result of higher metallicity galaxies being biased towards having higher stellar masses.  In the bottom panel of Figure~\ref{metal_effect2}, we show O32 versus mean HI ionisation rate inside the virial radius colour-coded by $f_{\rm esc}$.  As expected, we find that the galaxies with higher O32 are biased towards higher ionisation parameter.  However, there is considerable scatter due to the degeneracy with metallicity and ionisation parameter. While it is difficult to identify trends in O32-$f_{\rm esc}$ plane, there is a hint that the galaxies in our simulation with the highest O32 (e.g. ${\rm O32}>1$) have lower metallicities and higher HI ionisation rates.

In summary, the relationship between O32 and $f_{\rm esc}$ is complex and dictated by quantities such as metallicity, ionisation parameter, morphology, and feedback. The majority of the galaxies in our simulations with high $f_{\rm esc}$ are biased towards having high O32, consistent with observations \citep{Izotov2018b,Nakajima2019}. Quantitatively, the majority of our LyC leakers have ${\rm O32}> 0$ and all of the observed LyC leakers have O32 greater than this value.  Using ${\rm O32}= 0$ as a dividing line, if we normalise to the number of galaxies in both populations only 1\% of the galaxy population with ${\rm O32} < 0$ have $f_{\rm esc} > 10\%$.  In contrast 11\% of galaxies with ${\rm O32} > 0$  have $f_{\rm esc} > 10\%$.  Hence, galaxies with higher O32 are more biased to be LyC leakers. However, consistent with \cite{Bassett2019}, despite this bias, we do not find any other strong trends that may help identify more LyC leakers based on $f_{\rm esc}$ and O32 alone. Thus for isolated systems, high O32 may be a very common, but not sufficient condition for LyC leakage. It is important to note that \cite{Bassett2019} also identified two systems with high $f_{\rm esc}$ but low O32.  These two systems show evidence of merging and the shocks from this merger may decrease O32.  

\subsection{SII deficit}
Recently, observations from \cite{Wang2019} suggested that [SII] emission can be used to potentially identify LyC leakers at low redshift.  They demonstrated that LyC leakers exhibited systematically lower [SII] emission on the [SII] BPT diagram compared to typical, local star-forming galaxies. Hence, they argue that identifying systems with significant [SII] deficits is a useful tool for selecting those that have an ISM which is optically thin to ionising radiation.  

\begin{figure}
\centerline{\includegraphics[scale=1,trim={0 0.0cm 0 0.0cm},clip]{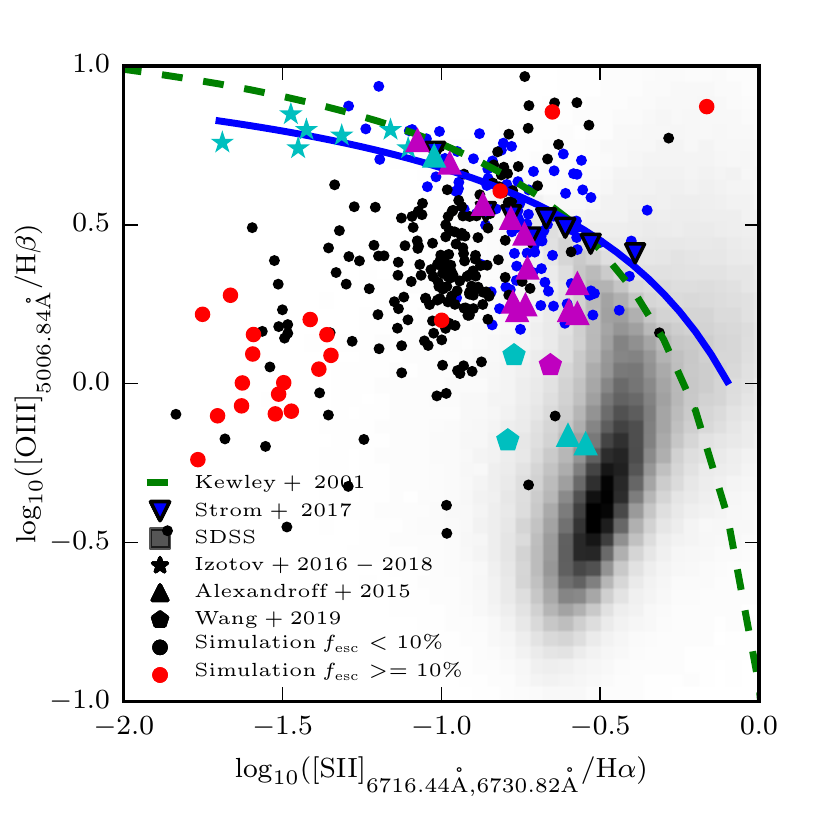}}
\caption{[SII] BPT diagram for simulated galaxies versus observations.  Local SDSS galaxies are shown as the grey 2D histogram.  The dashed green line shows the maximum theoretical starburst line from \protect\cite{Kewley2001}.  Blue points and the blue line and triangles represent observations of individual star-forming galaxies and the mean relation at $z=2-3$ from \protect\cite{Strom2017}.  Individual galaxies from our simulation are shown as circles.  Black circles represent systems that have $f_{\rm esc}<10\%$ while red points show galaxies in our simulations that are leaking a considerable amount of LyC radiation and have $f_{\rm esc}\geq10\%$.  Cyan stars represent leaky Green Pea Galaxies from \protect\cite{Izotov2018b,Izotov2018a,Izotov2016b,Izotov2016a}.  Cyan and magenta triangles represent leaky and non-leaky Lyman-Break analogues from \protect\cite{Alexandroff2015}.  Finally, cyan and magenta pentagons represent leaky and non-leaky star-forming galaxies from \protect\cite{Wang2019}. The regions of the {SII} BPT diagram populated by the simulated high-redshift leakers is clearly different from the low-redshift galaxies. The regions of this diagram populated by simulated leakers and non-leakers is clearly different indicating that if low-redshift systems are identified that overlap with our simulated high-redshift galaxies, this diagnostic could potentially be used to classify them as leakers.}
\label{SII_deficit}
\end{figure}

In Figure~\ref{SII_deficit}, we plot the [SII] BPT diagram for galaxies in our simulations compared to observations.  Our simulated high-redshift systems (i.e. with $12\geq z\geq9.2$) do show an offset on this diagram towards lower [SII]/H$\alpha$, more consistent with the expectations of $z\sim2-3$ galaxies (see \citealt{Katz2019b,Strom2017}). The black points on this plot represent simulated systems with $f_{\rm esc}<10\%$ while the red points represent the simulated systems with $f_{\rm esc}\geq10\%$.  We find that the majority of leakers have a systematically lower [SII]/H$\alpha$ than the non-leakers; however, these systems also tend to have systematically lower [OIII]/H$\beta$ as well, which was not observed by \cite{Wang2019}.  This behaviour is indicative of low metallicity (which for our systems ranges from $10^{-4}\leq Z/Z_{\odot}\leq0.4$).  From Figure~\ref{global_props}, there is a strong trend between $f_{\rm esc}$ and mass such that the lower mass haloes tend to have higher $f_{\rm esc}$.  These lower mass haloes also tend to have lower metallicity and thus the trend that we see in Figure~\ref{SII_deficit} is reflecting this effect. The reionization-era simulations of \cite{Barrow2017} also show numerous galaxies scattering to the bottom left of the [SII] BPT diagram, consistent with this being a metallicity effect.

The [SII]-deficient LyC leakers presented in \cite{Wang2019,Alexandroff2015,Izotov2016a,Izotov2018b} do not overlap the regions of the [SII] BPT diagram that are populated by our simulated high-redshift LyC leakers and thus we can conclude that these observed, low-redshift [SII]-deficient galaxies may not necessarily be analogues of the galaxies that reionized the Universe. Nevertheless, it is possible that low-redshift galaxies will be found that populate the same regions of the [SII] BPT diagram that are occupied by our simulated-high-redshift galaxies.  The simulated LyC leakers tend to be offset from the non-leakers in this diagram suggesting that it could still be a useful diagnostic for identifying LyC leakers.  The most ideal diagnostics that we derive from our simulations are further discussed in the next section.

\begin{figure}
\centerline{\includegraphics[scale=1,trim={0 0.0cm 0 0.0cm},clip]{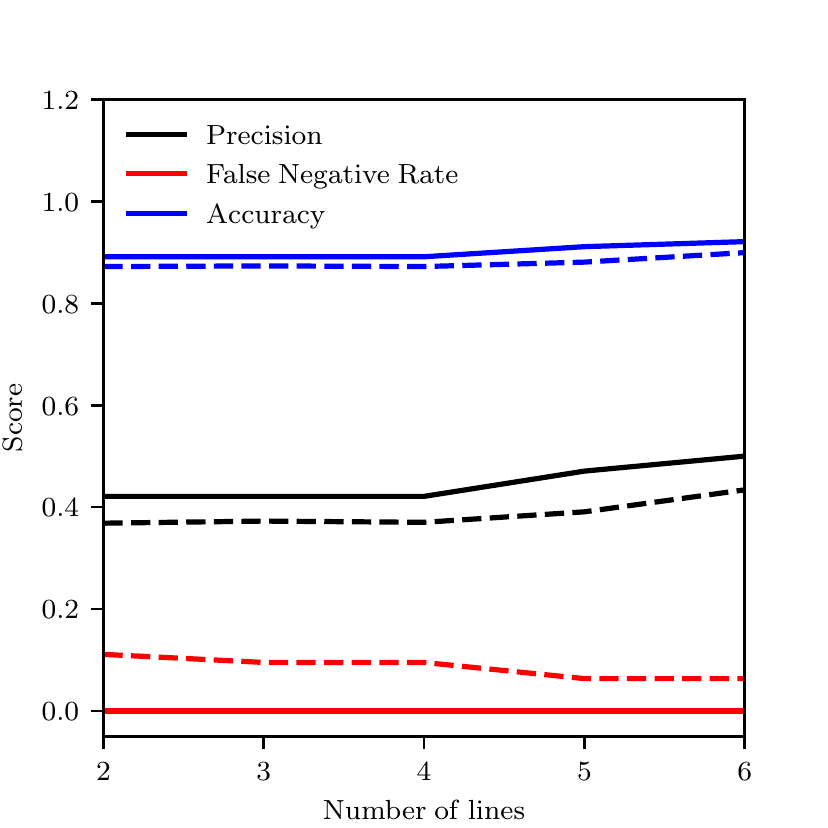}}
\caption{Accuracy (blue), precision (black), and false negative rate (red) as a function of the number of lines in the model for the training set (dashed) and test set (solid).  These metrics are fairly consistent between the training and test sets indicating that the models generalise well.  Both the precision and accuracy increase with the number of lines used in the model while the false negative rate remains both low and constant.}
\label{obs_predict}
\end{figure}

\begin{figure*}
\centerline{\includegraphics[scale=1,trim={0 0.0cm 0 0.0cm},clip]{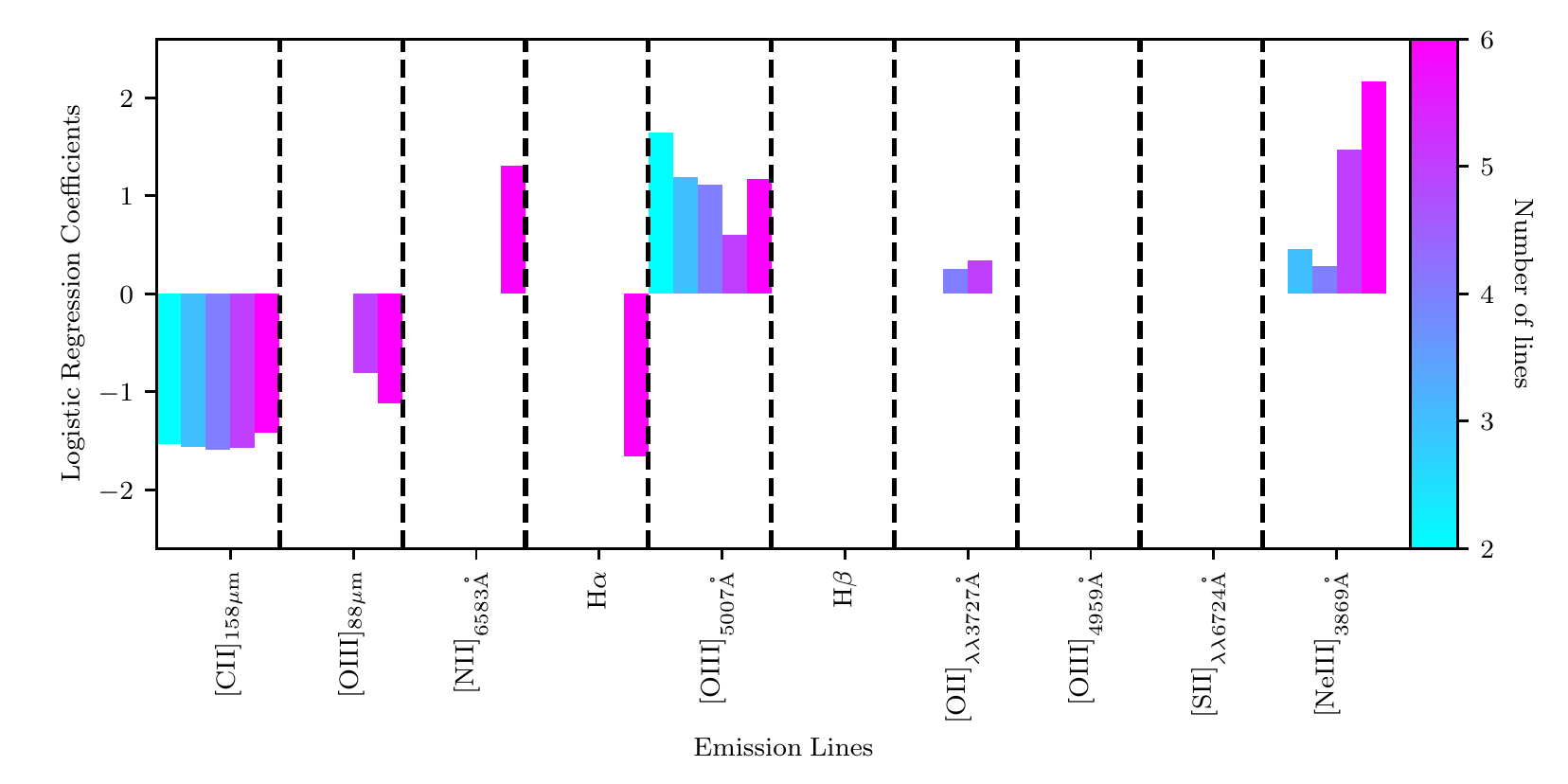}}
\caption{Logistic regression coefficients fitted for the different models. Negative values indicate that $f_{\rm esc}$ anti-correlates with the strength of the emission line while positive values show lines that have a positive correlation with $f_{\rm esc}$.  A zero coefficient represents a line that is not used for the prediction. As the regularisation is decreased, the number of lines included in the model increases as given by the colour of the bar.}
\label{logit_coeff}
\end{figure*}

\section{New Observational Diagnostic to Select LyC Leakers}
\label{diagnostic}

In this Section, we use our simulations to try and develop new methods for selecting LyC leakers from observations. We test these new methods against a sample of dwarf galaxies from the local Universe to identify potential local LyC leakers as well as a sample from $z>6$ to identify potential galaxies that actively played a role in reionizing the Universe. 

\subsection{Multi-band Models}
Because we have nearly complete information for each of our simulated galaxies on both nebular and infrared emission line luminosities as well as LyC escape fractions, we aim to develop a diagnostic that will most efficiently differentiate leakers from non-leakers.  To do this, we combine sparse modelling with simple machine learning algorithms to efficiently select the emission line combinations that best correlate with $f_{\rm esc}$. For this experiment, we use the same sample as above and only include haloes that are sampled by at least 300 dark matter particles that contain at least 20 star particles.

Our method is as follows: We set up a feature matrix containing the logarithm of the emission line luminosities of ${\rm [CII]_{158\mu m}}$, ${\rm [OIII]_{88\mu m}}$, ${\rm [NII]_{6583\angstrom}}$, ${\rm H\alpha}$, ${\rm [OIII]_{5007\angstrom}}$, ${\rm H\beta}$, ${\rm [OII]_{\lambda\lambda 3727\angstrom}}$, ${\rm [OIII]_{4959\angstrom}}$, ${\rm [SII]_{\lambda\lambda 6724\angstrom}}$, and ${\rm [NeIII]_{3869\angstrom}}$ for each galaxy in our simulation across all simulation snapshots\footnote{Note that for some galaxies, we predict very low luminosities for the emission lines.  This occurs mainly in the haloes with very low mass.  In order for this not to impact our training in this work, we set a minimum luminosity of 10$L_{\odot}$ for each emission line in this section.}.  The IR lines were selected for this exercise as they represent some of the brightest IR emission lines and are the two commonly observed IR lines in the epoch of reionization. The optical lines were chosen because they appear in common strong line diagnostics. Each galaxy is assigned a status of leaker or non-leaker depending on whether $f_{\rm esc}$ is greater than or less than 10\%. We then perform a stratified split of the data into a training set and a test set with a ratio of 80\%:20\% so that each data set has an equal fraction of leakers to non-leakers.  The latter set is to determine how well the model generalises to unseen data.  

Once the data has been prepared, we train a logistic regression model to best classify galaxies in the training set as either leakers or non-leakers.  This is done by minimising the binary cross-entropy subject to a weighting-scheme where we set class weights proportional to $1-\frac{N_{\rm class}}{N_{\rm galaxies}}$, where $N_{\rm class}$ is the total number of galaxies in each class.  We apply this weighting scheme because there is a high imbalance in the number of leakers to non-leakers in our data set and by weighting our cost-function, we can ensure that the scheme is not biased towards predicting more non-leakers correctly.

For practical purposes, we would like to build the best classifier that minimises the number of lines necessary to make the prediction as the more lines required for the prediction increases the expense of observations.  For this reason, we train multiple models where we apply an L1 regularisation scheme with different regularisation strengths.  We use the training set to search a grid of inverse regularisation strengths between 0.1 and 10 at intervals of 0.05 and we select the different models such that they have the greatest inverse regularisation strength per number of lines. Once we select out the lines that are most informative, we retrain the model without regularisation to find the best fit model for that set of lines. In principle, adding lines to the model will make it more accurate while decreasing the inverse regularisation strength will decrease the accuracy of the model.  By sampling multiple models, that use different numbers of lines for the prediction, we can balance the cost of the observation with accuracy.

In Figure~\ref{obs_predict} we show the accuracy, false negative rate, and precision of the models as a function of the number of lines used in the classifier.  For all models, these values are relatively consistent between the training set (dashed lines) and the test set (solid lines) indicating that the models generalise very well.  As expected, if we increase the number of lines used for classification, the accuracy (i.e. the fraction of galaxies that we correctly classify as leakers or non-leakers) increases.  Likewise, the precision (i.e. the fraction of galaxies that the classifier predicts as leakers that actually are leakers) also increases.  When the number of lines used for classification increases from two to six, the precision increases from $\sim40\%$ to $\sim50\%$.  This value is arguably the most important as it indicates that if we applied our model to real data and then observed the galaxies with a high leaker probability, only 40\%-50\% would turn out to be true leakers.  Given the difficulty in observing leakers in the local Universe, this high predicted hit rate is very promising. In contrast, the false negative rate (i.e. the fraction of leakers that we misclassify as non-leakers) remains very low ($\lesssim10\%$) for all models. This is due to the fact that we artificially increased the class weights of the leakers when training the model so as to not misclassify them.  This certainly impacts the total accuracy of the model and decreases it compared to one where we do not adjust the class weights; however, incorrectly classifying true leakers is arguably worse than having a slightly lower precision and accuracy.

By analysing the coefficients in the trained logistic regression model, we can better understand how different emission lines correlate with the LyC escape fraction and which lines can be used to classify leakers and non-leakers. Note that in order for this to work, emission line strengths have to be properly normalised and when training the models, we performed a standard scaling by subtracting the mean and dividing by the standard deviation. In Figure~\ref{logit_coeff}, we show the logistic regression coefficients for each emission line for the different models.  The only non-zero coefficients in our two-line model are those of the ${\rm [CII]_{158\mu m}}$ and the ${\rm [OIII]_{5007\angstrom}}$ lines.  Because the coefficients are negative and positive, respectively, this indicates that $f_{\rm esc}$ negatively correlates with ${\rm [CII]_{158\mu m}}$ and positively correlated with ${\rm [OIII]_{5007\angstrom}}$. This is expected from our simulations because we know that [CII] emission is highly correlated with the presence of neutral gas \citep{Katz2017,Katz2019b} while the higher ionisation line [OIII] is much more representative of ionised gas and SN feedback \citep{Katz2019b}. However, it is important to note that in the nearby Universe, for higher metallicity systems, a substantial, albeit still subdominant fraction of the [CII] emission is expected to originate from ionised regions (see Figure~9 of \citealt{Cormier2019}).

\begin{figure}
\centerline{\includegraphics[scale=1]{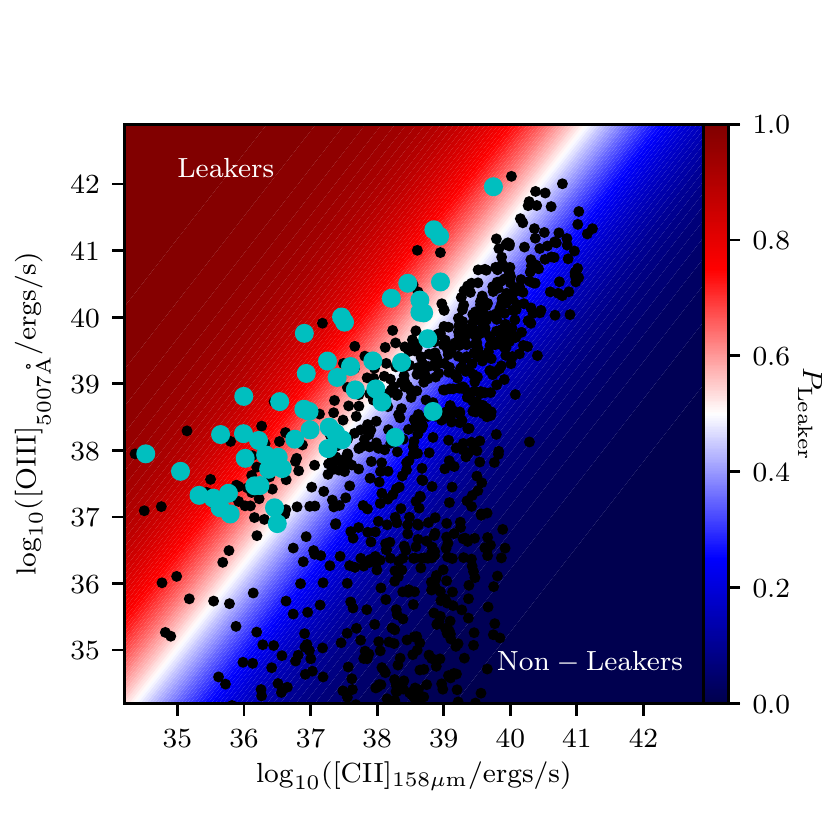}}
\caption{${\rm [CII]_{158\mu m}}$ versus ${\rm [OIII]_{5007\angstrom}}$ for our simulated galaxies.  The black points represent galaxies with $f_{\rm esc}<10\%$ while the cyan points represent galaxies with  $f_{\rm esc}\geq10\%$. The diverging colour map that is underlaid represents the probability that a galaxy is a leaker or a non-leaker from our most simple, two-line logistic regression model. The white band represents the dividing line for when a galaxy is classified as a leaker or a non-leaker in our model. Nearly all of the simulated leakers fall to the left of this line demonstrating that the model has high sensitivity. }
\label{lr_map}
\end{figure}

In Figure~\ref{lr_map} we show how the two-line model separates the space of ${\rm [CII]_{158\mu m}}$ and  ${\rm [OIII]_{5007\angstrom}}$ into regions where it predicts an $f_{\rm esc}\geq10\%$ and $f_{\rm esc}<10\%$.  The cyan and black points represent leakers and non-leakers in the simulation, respectively while the red and blue regions show where the model predicts high and low $f_{\rm esc}$, respectively. The white region represents the dividing line of being classified as a potential leaker by our model.  Note that because of the class-weighting scheme we applied, this does not represent an actual probability.  The true probability has to be calculated from the results from the test set that show a probability of 40\% of being a leaker in the red region (as given by the positive predictive value of the model). This is most easily visualised by looking at the amount of contamination of black points in the red region. Overall, it is clear from Figure~\ref{lr_map} that the logistic regression model is able to efficiently separate the true leakers from the bulk of the galaxy population. 

As the model becomes more complex (due to an increasing inverse regularisation strength), other lines are included in the model.  After ${\rm [OIII]_{5007\angstrom}}$ and ${\rm [CII]_{158\mu m}}$, the next most important line for classifying a galaxy as a leaker or a non-leaker is the ${\rm [NeIII]_{3869\angstrom}}$ line. For our most complex models, this line has a coefficient with a similar strength to the ${\rm [OIII]_{5007\angstrom}}$ line.  The coefficient for ${\rm [NeIII]_{3869\angstrom}}$ is also positive indicating that it correlates with a galaxy being a leaker.  This is expected because ${\rm [NeIII]_{3869\angstrom}}$ often traces ${\rm [OIII]_{5007\angstrom}}$ although it has a slightly higher ionisation potential and a much higher critical density.  Interestingly, as the model becomes more complex, the  ${\rm [OIII]_{88\mu m}}$ line becomes important but unexpectedly, the coefficient is negative. This line is emitted from gas at much higher densities compared to ${\rm [OIII]_{5007\angstrom}}$ and is perhaps balancing the contribution of the ${\rm [OIII]_{5007\angstrom}}$ line to the prediction by excluding any contribution from the higher density gas. Interestingly, in no model is the ${\rm [SII]_{\lambda\lambda6724\angstrom}}$ doublet selected as a good indicator for identifying leakers from non-leakers, nor is ${\rm [OIII]_{4959\angstrom}}$ which contains the same information as ${\rm [OIII]_{5007\angstrom}}$ in terms of it originating from the same regions of the density-temperature phase space.  Theoretically, the line ratio of ${\rm [OIII]_{5007\angstrom}}/{\rm [OIII]_{4959\angstrom}}$ should be three \citep{Storey2000}. What seems to matter most for whether a line is a good indicator of $f_{\rm esc}$ (or whether it provides any additional information compared to other lines) is which regions of the density-temperature phase space that it originates.

To classify an observed galaxy is a leaker or non-leaker, our models can be employed using the following equation:
\begin{equation}
    P(f_{\rm esc}\geq10\%)=\frac{e^{\Sigma_{i=0}^{10}C_iX_i}}{e^{\Sigma_{i=0}^{10}C_iX_i}+1},
\end{equation}
where the values of the coefficients, $C_i$, are given for the minimal and complete models in Table~\ref{lr_coef_tab}. $X_i$ represents the $\log_{10}$ of a given line luminosity in solar units normalized by our standard scaling. The line that corresponds to each coefficient is also given in Table~\ref{lr_coef_tab}.  $C_0$ is not associated with any line as it represents the intercept in the model and thus $X_0=1$.

\begin{table*}
    \centering
    \begin{tabular}{llcccccccc}
    \hline
    $C_i$ & $X_i$ & $\mu_i$ & $\sigma_i$ & ${\rm N_{lines}=2}$ & ${\rm N_{lines}=3}$ & ${\rm N_{lines}=4}$ & ${\rm N_{lines}=5}$ & ${\rm N_{lines}=6}$  & IR-Only Model\\
    \hline
        $C_0$ & 1 & - & - & -1.27 & -1.28 & -1.26 & -1.33 & -1.88 & -0.49\\
        $C_1$ & $\frac{\log_{10}({\rm [CII]_{158\mu m}})-\mu_1}{\sigma_1 L_{\odot}}$ & 38.24 &  1.60 & -1.53 & -1.56 & -1.59 & -1.57 & -1.41 & -0.98 \\
        $C_2$ & $\frac{\log_{10}({\rm [OIII]_{88\mu m}})-\mu_2}{\sigma_2 L_{\odot}}$ & 36.14 & 2.14 & 0 & 0 & 0 & -0.81 & -1.12 & 0.59 \\
        $C_3$ & $\frac{\log_{10}({\rm [NII]_{6583\angstrom}})-\mu_3}{\sigma_3L_{\odot}}$ & 36.91 & 1.84 & 0 & 0 & 0 & 0 & 1.31 & 0 \\
        $C_4$ & $\frac{\log_{10}({\rm H\alpha})-\mu_4}{\sigma_4L_{\odot}}$ & 38.79 & 1.32 & 0 & 0 & 0 & 0 & -1.66  & 0 \\
        $C_5$ & $\frac{\log_{10}({\rm [OIII]_{5007\angstrom}})-\mu_5}{\sigma_5L_{\odot}}$ & 37.47 & 2.21 & 1.65 & 1.19 & 1.11 & 0.60 & 1.18  & 0 \\
        $C_6$ & $\frac{\log_{10}({\rm H\beta})-\mu_6}{\sigma_6L_{\odot}}$ &  38.16 & 1.31 & 0 & 0 & 0 & 0 & 0  & 0 \\
        $C_7$ & $\frac{\log_{10}({\rm [OII]_{\lambda\lambda 3727\angstrom}})-\mu_7}{\sigma_7L_{\odot}}$ & 37.53 & 1.75 & 0 & 0 & 0.26 & 0.34 & 0  & 0 \\
        $C_8$ & $\frac{\log_{10}({\rm [OIII]_{4959\angstrom}})-\mu_8}{\sigma_8L_{\odot}}$ & 37.03 & 2.13 & 0 & 0 & 0 & 0 & 0  & 0 \\
        $C_9$ & $\frac{\log_{10}({\rm [SII]_{\lambda\lambda 6724\angstrom}})-\mu_9}{\sigma_9L_{\odot}}$ & 37.46 & 1.61 & 0 & 0 & 0 & 0 & 0  & 0  \\
        $C_{10}$ & $\frac{\log_{10}({\rm [NeIII]_{3869\angstrom}})-\mu_{10}}{\sigma_{10}L_{\odot}}$ & 36.51 & 1.86 & 0 & 0.46& 0.29& 1.47 & 2.17  & 0 \\
    \hline
    \end{tabular}
    \caption{Coefficient values and the associated lines for the different models.  These should be input into the logistic regression probability equation to classify galaxies as LyC leakers or not. }
    \label{lr_coef_tab}
\end{table*}

\subsection{IR-only Model}
\subsubsection{Model Development}
While potentially very useful in identifying LyC leakers, one of the main downsides to the above diagnostics is that they use a mix between IR and nebular lines.  Because there are no instruments that can simultaneously sample this huge range in frequency, observations from multiple instruments will have to be combined. This in itself can cause issues due to the systematic differences between the various telescopes and there are simply far fewer galaxies that have been observed across such a broad frequency range. For this reason, we develop a second diagnostic that only relies on IR lines; in particular, the ${\rm [CII]_{158\mu m}}$ and ${\rm [OIII]_{88\mu m}}$ emission lines, which have already been suggested to be good indicators of $f_{\rm esc}$ \citep{Inoue2016}. One of the primary advantages of such a model is that the combination of these two lines has already been constrained out to $z=9.1$ \citep{Laporte2019} while in the local Universe, larger samples of dwarf galaxies exist for which the luminosity of each of these lines has been measured \citep[e.g.][]{Madden2013}.

\begin{figure}
\centerline{\includegraphics[scale=1]{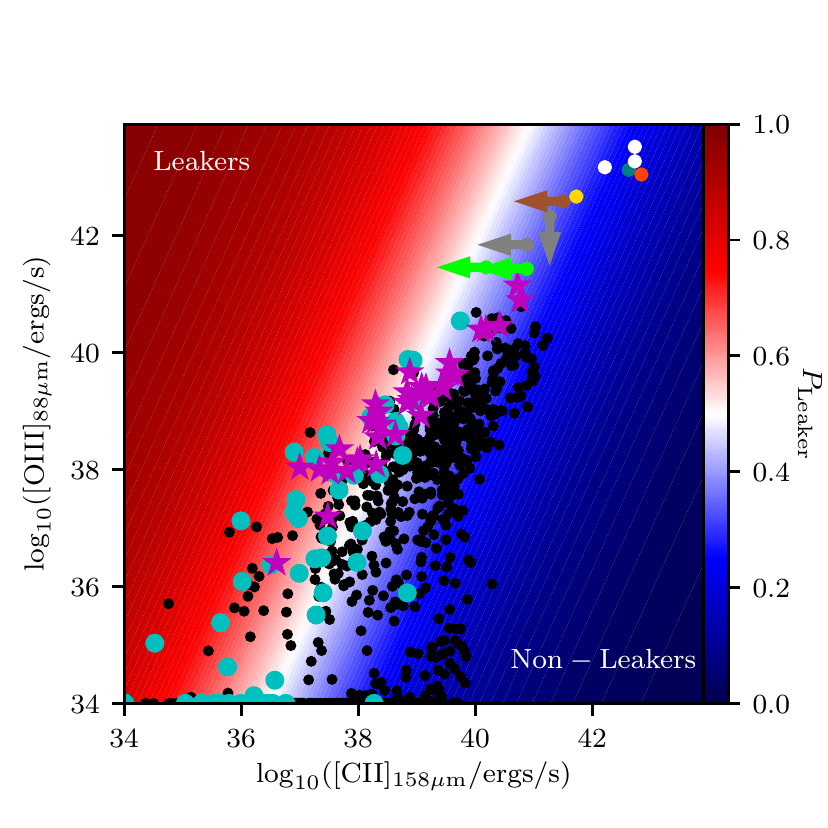}}
\caption{${\rm [CII]_{158\mu m}}$ versus ${\rm [OIII]_{88\mu m}}$ for our simulated galaxies.  The black points represent galaxies with $f_{\rm esc}<10\%$ while the cyan points represent galaxies with  $f_{\rm esc}\geq10\%$. The diverging colour map that is underlaid represents the probability that a galaxy is classified as a leaker or a non-leaker from our IR logistic regression model. The white band represents the dividing line for when a galaxy is classified as a leaker or a non-leaker in our model. Low-redshift observations of local dwarf galaxies are shown as the magenta stars \protect\citep{Madden2013,Cormier2015}. High-redshift observations of galaxies in the epoch of reionization are shown in lime \protect\citep{Laporte2019}, teal \protect\citep{Hashimoto2019}, grey \protect\citep{Carniani2017}, gold \protect\citep{Bakx2020}, orange-red \citep[companion galaxy only,][]{Walter2018}, brown \protect\citep{Inoue2016}, and white \protect\citep{Harikane2019}. Arrows represent upper limits. Note that for the \protect\cite{Carniani2017} observations, we have treated the [CII] clump separately from the [OIII] clump and optical clump on this plot as otherwise, the limits from non-detections would dominate. Our conclusions about this object would not change if we treated both clumps as one system.} 
\label{lr_IR_map}
\end{figure}

The downside of this IR-model is that it will be less accurate than the two-line model presented above. This is because if it was the optimal model, with the regularisation scheme applied above, it would have been naturally selected. Moreover, as described earlier, the ${\rm [OIII]_{88\mu m}}$ emission line coefficient from the earlier models was always negative. To first order, one might have assumed this meant that $f_{\rm esc}$ negatively correlates with ${\rm [OIII]_{88\mu m}}$, in which case, having a model that combines ${\rm [CII]_{158\mu m}}$ and ${\rm [OIII]_{88\mu m}}$ makes little sense. Rather, our interpretation is that the peak of ${\rm [OIII]_{88\mu m}}$ emission occurs at slightly higher densities than ${\rm [OIII]_{5007\angstrom}}$ emission and thus by having a negative coefficient, the model was removing the higher density part of phase-space where OIII can form. If this interpretation is correct, we should still find success using ${\rm [CII]_{158\mu m}}$ and ${\rm [OIII]_{88\mu m}}$ because ${\rm [OIII]_{88\mu m}}$ is sensitive to the feedback in the galaxy, but it will be less optimal than using ${\rm [OIII]_{5007\angstrom}}$ because of the difference in gas density where the emission peaks.

The behaviour of the ${\rm [OIII]_{88\mu m}}$ compared to ${\rm [OIII]_{5007\angstrom}}$ may seem curious in our simulation because ${\rm [OIII]_{5007\angstrom}}$ has a considerably higher critical density ($7\times10^{5}{\rm cm^{-3}}$ \citealt{Baskin2005}) compared to ${\rm [OIII]_{88\mu m}}$ ($\sim500{\rm cm^{-3}}$ \citealt{Spinoglio2012}). Therefore, one may expect that the line ratio of ${\rm [OIII]_{5007\angstrom}}$ to ${\rm [OIII]_{88\mu m}}$ should increase with density \citep[e.g.][]{Keenan1990} as collisional de-excitaton should reduce the scaling of the ${\rm [OIII]_{88\mu m}}$ line with density above its critical density compared to ${\rm [OIII]_{5007\angstrom}}$.  In contrast, our simulations show that the ratio actually increases with density (see Figure~3 of \citealt{Katz2019b}). This occurs because the regions of our simulation that are bright in both ${\rm [OIII]_{5007\angstrom}}$ and ${\rm [OIII]_{88\mu m}}$ rarely reach the critical density of ${\rm [OIII]_{88\mu m}}$; hence we do not probe the expected drop off at high densities due to collisional de-excitation.  Furthermore, the CMB at $z=10$ reduces the emergent line flux of ${\rm [OIII]_{88\mu m}}$, particularly at lower densities, while having no impact on ${\rm [OIII]_{5007\angstrom}}$. This explains the behaviour of why ${\rm [OIII]_{88\mu m}}$ peaks at higher densities compared to ${\rm [OIII]_{5007\angstrom}}$ in our simulation.  We have confirmed that by rerunning the {\small CLOUDY} modules with a $z=0$ CMB background, we recover the expected behaviour.

Fortunately, the best model using only ${\rm [CII]_{158\mu m}}$ and ${\rm [OIII]_{88\mu m}}$ performs only slightly worse than the  ${\rm [CII]_{158\mu m}}$ and ${\rm [OIII]_{5007\angstrom}}$ model.  The parameters for this model can be found in Table~\ref{lr_coef_tab}.  We find that the IR-only model has an accuracy of 86\%, a precision of 36\% and a false negative rate of 6.3\% measured on the test set.  Both the accuracy and precision have slightly decreased compared to the optimal model while the false negative rate has increased; however, overall, the performance is not significantly worse. In Figure~\ref{lr_IR_map}, we plot ${\rm [CII]_{158\mu m}}$ versus ${\rm [OIII]_{88\mu m}}$ for all galaxies in our simulation.  LyC leakers that have an $f_{\rm esc}\geq10\%$ are shown in cyan while simulated galaxies with $f_{\rm esc}<10\%$ are shown in black.  Compared to Figure~\ref{lr_map}, one can see that there is less differentiation between leakers and non-leakers in the  ${\rm [CII]_{158\mu m}}$-${\rm [OIII]_{88\mu m}}$ space compared to the ${\rm [CII]_{158\mu m}}$-${\rm [OIII]_{5007\angstrom}}$ space. The underlaid colour map in Figure~\ref{lr_IR_map} represents the probability given by the logistic regression model that an individual system will have an $f_{\rm esc}\geq10\%$. Systems that fall close to the white band have approximately an equal probability of being classified as a LyC leaker. Qualitatively, it is easy to see that the model does a good job in separating leakers from non-leakers.

\subsubsection{Comparison with Observations}
The IR-only model was built in the context of comparing to observations from both low and high redshift. In Figure~\ref{lr_IR_map}, we plot 37 galaxies from the local Dwarf Galaxy Survey \citep[DGS,][]{Madden2013} for which ${\rm [CII]_{158\mu m}}$ and ${\rm [OIII]_{88\mu m}}$ have been measured \citep{Cormier2015} as magenta stars. Interestingly, the local galaxy sample straddles the dividing line between where our model predicts galaxies which have an $f_{\rm esc}$ greater than or less than 10\%, allowing us to determine how well it performs on real galaxies.

Unfortunately, observational escape fraction measurements are not available for most DGS galaxies.  Thus it is difficult to compare whether the accuracy, precision, and false negative rate of our model coincides with reality. Instead we discuss a few notable galaxies and how our predictions compare with observations.

Of the galaxies predicted to be leakers by our model, NGC 2366 has one of the highest probabilities of having an $f_{\rm esc}\geq10\%$.  NGC~2366 is the closest known analogue of a Green Pea galaxy, many of which have already been determined to be LyC leakers \citep{Izotov2016a,Izotov2016b,Izotov2018a,Izotov2018b}. While an escape fraction has not yet been measured for this system, IFU observations revealed strong ionised outflows that are almost certainly optically thin and can facilitate the escape of LyC radiation \citep{Micheva2019}. NGC~4861 is another galaxy that our model predicts to be a leaker, although it is less confident about this galaxy compared to NGC~2366. NGC~4861 is similar to NGC~2366 both visually and the fact that it contains an outflow likely driven by supernova feedback \citep{vEymeren2009}. Once again, no escape fraction has been measured for this galaxy but it is known from simulations that strong stellar feedback is needed to facilitate the escape of ionising photons and it is encouraging that our model has selected these two galaxies, both of which show evidence for being disrupted by feedback. NGC~1705 is another example of a galaxy selected to be a leaker by our model.  It has a similar probability to that of NGC~4861.  Once again, in this system, there is clear evidence that the ISM is being disrupted by stellar feedback; however, detailed observations demonstrate that the outflow is likely optically thick \citep{Zastrow2013}.

One of the more well studied galaxies in the context of escaping LyC photons is Haro~11.  Our model is fairly confident that Haro~11 has $f_{\rm esc}<10\%$, assigning it a 22\% probability of having $f_{\rm esc}>10\%$.  Interestingly, LyC radiation has been directly observed for this system and the estimate is that $f_{\rm esc}=3.3\%$ \citep{Leitet2011}, confirming the prediction from our model. The model similarly predicts minimal Lyman continuum leakage in both Haro~2 and also Haro~3.

Finally, our model predicts two other systems as sitting right on the boundary between having $f_{\rm esc}\geq10\%$.  These two systems are Tol~1214-277 and NGC~5253.  In the case of Tol~1214-277, Ly$\alpha$ observations are highly suggestive that there might be LyC leakage \citep{Verhamme2015} and for NGC~5253, a dwarf starburst galaxy, there is an ionisation cone that suggests LyC leakage as well, but it is not clear how much \citep{Zastrow2011}.

Our model is of course not perfect as given by the fact that the false negative rate is non-zero and the accuracy is not 100\%.  For example, we predict that SBS~0335-052 has $f_{\rm esc}\geq10\%$, however, the observations from \cite{Grimes2009} show no evidence of emission beyond the Lyman limit. It should be noted that this work also did not infer the detection of LyC emission from Haro~11, which is contradictory to the conclusions from \cite{Leitet2011} and viewing angle will of course play a role. Nevertheless, it must be noted that we predict a contamination rate of up to 65\% among the candidates we predict as leakers so some incorrect predictions are expected. A summary of all systems that we make predictions for can be found in Table~\ref{fesc_pred}.

\begin{table*}
    \centering
    \begin{tabular}{lcp{12cm}}
	\hline
	Galaxy & $P(f_{\rm esc}\geq10\%)$ & Notes \\
	\hline
	Haro 11 & 0.22 &$f_{\rm esc}=3.3\pm0.7\%$ \citep{Leitet2011}. \\
	Haro 2 & 0.34 & -\\
	Haro 3 & 0.36 & $f_{\rm esc}<0.1\%$ based on CII \citep{Grimes2009}. \\
	He 2-10 & 0.34 & Confirmed to host an AGN.  Observations suggest that the galaxy is optically thick; however, there may be small ionised channels where radiation escapes \citep{Zastrow2013}.\\
	HS 0017+1055 & 0.57 & -\\
	HS 0052+2536  & 0.27 & -\\
	HS 1222+3741 & 0.54 & -\\
	HS 1304+3529  & 0.40 & -\\
	HS 1319+3224 & 0.61 & -\\
	HS 1330+3651  & 0.39 & -\\
	II Zw 40 & 0.39 & Emission is likely to be shock induced and the galaxy is unlikely to be optically thin \citep{Jaskot2013}.\\
	I Zw 18 & 0.71 & -\\
	Mrk 1089 & 0.22 & -\\
	Mrk 1450 & 0.61 & -\\
	Mrk 153  & 0.47 & -\\
	Mrk 209 & 0.76 & -\\
	Mrk 930 & 0.29 & -\\
	NGC 1140 & 0.31 & $f_{\rm esc}<1.4\%$ based on CII \citep{Grimes2009}. \\
	NGC 1569 & 0.45 & -\\
	NGC 1705 & 0.61 & Clear evidence that the ISM is being disrupted by stellar feedback; however, detailed observations demonstrate that the outflow is likely optically thick \citep{Zastrow2013}.\\
	NGC 2366 & 0.72 & The closest known analogue of a Green Pea galaxy. IFU observations revealed strong ionised outflows that are almost certainly optically thin and can facilitate the escape of LyC radiation \citep{Micheva2019}.\\
	NGC 4214 & 0.47 & Some of the UV radiation may be escaping based on the dust SED \citep{Hermelo2013}.\\
	NGC 4861 & 0.63 & Similar to NGC~2366 both visually and the fact that it contains an outflow likely driven by supernova feedback \citep{vEymeren2009}. \\
	NGC 5253 & 0.49 & There is an ionisation cone that suggests LyC leakage, but it is not clear how much \citep{Zastrow2011}.\\
	NGC 625 & 0.56 & -\\
	Pox 186  & 0.83 & -\\
	SBS 0335-052 & 0.64 & No evidence of emission beyond the Lyman limit \cite{Grimes2009}.\\
	SBS 1159+545 & 0.60 & -\\
	SBS 1415+437 & 0.64 & -\\
	SBS 1533+574 & 0.42 & -\\
	Tol 1214-277 & 0.51 &  Ly$\alpha$ observations are highly suggestive that there might be LyC leakage \citep{Verhamme2015}.\\
	UGC 4483  & 0.76 & -\\
	UM 133  & 0.50 & -\\
	UM 311  & 0.35 & -\\
	UM 448  & 0.19 & -\\
	UM 461  & 0.72 & -\\
	VII Zw 403  & 0.63 & -\\
	\hline
    \end{tabular}
    \caption{Probability that our model classifies a galaxy of having an $f_{\rm esc}\geq10\%$ for galaxies from the Dwarf Galaxy Survey \citep{Madden2013} as given by our IR-only model.}
    \label{fesc_pred}
\end{table*}

Since ALMA has come fully online, samples of high-redshift galaxies with observations of both ${\rm [CII]_{158\mu m}}$ and ${\rm [OIII]_{88\mu m}}$ have been made up to $z=9.1$ \citep[e.g.][]{Laporte2019}. Thus our IR-only model can be used to assess whether or not the galaxies that have been observed directly in the epoch of reionization are potentially the sources of reionization.  In Figure~\ref{lr_IR_map}, we plot measurements or constraints on ${\rm [CII]_{158\mu m}}$ and ${\rm [OIII]_{88\mu m}}$ for eleven galaxies at $z>6$ in lime \citep{Laporte2019}, teal \citep{Hashimoto2019}, grey \citep{Carniani2017}, gold \citep{Bakx2020}, orange-red \citep{Walter2018}, and brown \citep{Inoue2016}. Most of these systems have emission line luminosities that are significantly higher than the local dwarfs, which is not surprising because they have been observed at very high redshift. Because these luminosities extend beyond the galaxies in our sample, we have to trust that our model extrapolation suffices in this luminosity regime. Assuming that our model extrapolation holds, the two galaxies with definite ${\rm [CII]_{158\mu m}}$ measurements fall in the blue region of Figure~\ref{lr_IR_map}, indicating that they are unlikely to be leaking a significant amount of LyC photons into the IGM. In contrast, those systems for which we have only upper limits ${\rm [CII]_{158\mu m}}$ emission are still strong candidates to be LyC leakers.  

In particular, MACS1149\_JD1 from \cite{Laporte2019} sits very close to the boundary. The ${\rm [CII]_{158\mu m}}$ emission line luminosity that is shown represents a $3\sigma$ upper limit, indicating that it could be far less luminous.  If this is confirmed to be the case, based on our modelling, MACS1149\_JD1 represents a prime candidate for a galaxy that actively contributed to the reionization of the Universe.  We note that the actual luminosities for the ${\rm [CII]_{158\mu m}}$ and ${\rm [OIII]_{88\mu m}}$ emission lines of MACS1149\_JD1 and A2744\_YD4 are formally dependent on the amount of magnification which was assumed to be factors of 10 and 2 for each galaxy, respectively. Because the dividing line between being classified as a leaker and non-leaker in our model is not parallel to the one-to-one line, our conclusions about MACS1149\_JD1 and A2744\_YD4 will change if the actual magnification is much different from what is assumed. Furthermore, we note that a recent reanalysis of ALMA data for MACS1149\_JD1 and A2744\_YD4 has revealed [CII] detections that are more spatially extended compared to the [OIII] \citep{Carniani2020}.  These results would place these galaxies much closer the the low-redshift [CII]-SFR relation, consistent with the predictions from our simulations \citep{Katz2019b}; however, it would move the galaxies to the right on the [CII]-[OIII] diagram, decreasing their probability of being LyC leakers. In particular, this alternative [CII] measurement for MACS1149\_JD1 is $3\times$ higher than the previous upper limit (see Table~1 of \citealt{Carniani2020}).

\section{Discussion}
\label{discussion}

\subsection{Comparison with Previous Work}
Having developed new diagnostics to identify LyC leakers, it is important to consider how these compare to other methods. Earlier it was determined that the population of simulated galaxies was biased such that the fraction of galaxies that have ${\rm O32}>0$ that have a $f_{\rm esc}>10\%$ was approximately ten times greater than that with ${\rm O32}<0$. Indeed, the expected relations between O32 and $f_{\rm esc}$ from \cite{Faisst2016,Izotov2018a} rapidly rise beyond this value. These relations contradict those from \cite{Bassett2019} which demonstrate that the rapid rise in $f_{\rm esc}$ at high values of O32 is very sensitive to both the metallicity of the system as well as the ionisation parameter.  Our systems tend to have higher ionisation parameters and lower metallicities, consistent with systems closer to $z\sim3$ which favours a much more gradual increase in $f_{\rm esc}$ as a function of O32 (see the right panel of Figure~10 in \citealt{Bassett2019}).  Nevertheless, if we create a model to separate leakers from non-leakers based only on O32 and use ${\rm O32}=0$ as the dividing line, only one in ten systems that we follow up on would turn out to be true leakers. This performs significantly worse than our logistic regression model which has a precision of $40\%-50\%$. 

One can be more aggressive in their selection methods, for example, \cite{Izotov2016a,Izotov2016b,Izotov2018a} preferentially selected compact, low-metallicity star-forming galaxies that have  ${\rm O32}>0.69$.  With their selection criteria, they demonstrated that all six low-redshift galaxies in their sample had leaking LyC radiation. If we apply a cut in O32 at ${\rm O32}=0.69$, rather than 0, we find that our false negative rate increases dramatically from a few per cent to $>50\%$ because many of the simulated LyC leakers have slightly lower O32.  Hence this cut is not satisfying either.  However, in combination with another $f_{\rm esc}$ indicator, the false negative rate is expected to drop. 

Interestingly, our models put very little weight on the Balmer lines such as H$\alpha$ and H$\beta$.  This is potentially surprising given the fact that the strength of these lines is expected to be dominated by recombination physics and similarly, other models, such at that discussed in \cite{Zackrisson2013} partially base their estimate on H$\beta$. Our systems are not outliers in this regard as Figure~12 of \cite{Katz2019b} demonstrates that our simulated galaxies reasonably reproduce the expected Kennicutt relation between the Balmer lines and SFR. If one knew the intrinsic ionising photon production rate and the luminosity of H$\beta$, an estimate for $f_{\rm esc}$ could be made.  This method has been used in reverse to estimate the ionising photon production rate from known leakers \citep{Schaerer2016}. Given this exact coupling, it is perhaps curious as to why our model finds that these lines are not the best indicators for escaping LyC radiation. 

We believe that there are two clear distinctions that cause our machine learning models to prefer [OIII] and [CII].  First, we have defined our model to classify leakers from non-leakers based on a cut at $f_{\rm esc}=10\%$.  This is very different from trying to predict the exact value of $f_{\rm esc}$ from spectral properties.  In the latter situation, it is possible that a different method would be necessary that may be more akin to using the UV spectral slope and an equivalent width of a Balmer line \citep{Zackrisson2013}. This method is really only sensitive to $f_{\rm esc}\gtrsim50\%$.  In practice, it is still very difficult to determine the global $f_{\rm esc}$ from a galaxy observationally due to the strong variations in $f_{\rm esc}$ depending on sight-line. Second, it is well established from simulations that $f_{\rm esc}$ is strongly regulated by feedback \citep[e.g.][]{Kimm2014,Trebitsch2017,Kimm2017,Rosdahl2018,Ma2020}. Simulations show that SFR, which correlates very strongly with recombination lines \citep{Kennicutt1998}, is not a good indicator of $f_{\rm esc}$ \citep[e.g.][]{Yoo2020}. The emission lines most sensitive to the feedback, among the ones we have measured, are the [OIII] lines.  In contrast [CII] follows the neutral gas incredibly well.  A decrease in neutral gas favours high $f_{\rm esc}$ as does the presence of feedback and hence, our model selects the two lines that best track these quantities.

\subsection{Caveats}
While our simulations have been successful in modelling many of the properties of the observed high-redshift galaxy population, there are numerous caveats that should be kept in mind when interpreting our results.  

Like all simulations, the Aspen simulations have both finite spatial and mass resolution which can impact both the escape fraction and the simulated emission line luminosities. Significantly higher resolution will be needed in the ISM to better resolve the individual low-column density channels through which ionising photons escape \citep{Kimm2019,Kakiichi2019}.  Likewise, the resolution of our simulation just barely resolves a multiphase ISM which can significantly impact how emission line luminosities are modelled. For the less resolved galaxies, this can smooth over both the ISM and stellar distribution, both of which will impact our results.  A resolution study will be needed to confirm the quantitative results presented here. The simulation depends on a set of sub-grid feedback models that were designed so that our galaxy population matches an extrapolation of the high-redshift stellar mass-halo mass relation; however, this choice of feedback significantly impacts the state of the ISM as well as the LyC escape channels.  Furthermore, our simulations are not completely representative of the high-redshift galaxy population because we use a zoom-in simulation of a specific environment around a massive high-redshift galaxy. In order to achieve such high resolution for a massive galaxy system, we used a zoom-in technique rather than simulate the entire volume.  While this region contains more than 1,000 resolved galaxies, a full-box realisation will need to be completed in order to ensure that our results generalise. Extensions to simulations like SPHINX \citep{Rosdahl2018} or Renaissance \citep{Oshea2015} that probe the right mass range, redshift, and ISM physics could be ideal laboratories for similar experiments moving forward.

We re-emphasise that comparing to observations is non-trivial both because the LyC escape fractions are only observed along a single line-of-sight and our treatment of dust is fairly simplistic. In order to confirm or refute our model, larger samples of LyC leakers will be needed so that the comparison can be made in a probabilistically. Dust will not impact the IR emission line luminosities; however, IR emission line luminosities are sensitive to the exact yields we use for SN in our simulation which were based on solar abundances. It is known from observations that the C/O ratio evolves with metallicity \citep[e.g.][]{Pettini2008} as the production of C is expected to lag behind that of O. This could impact the galaxies from the DGS that we identify as leakers because it would shift all of the simulated galaxies by some amount to the left on Figure~\ref{lr_IR_map} resulting in fewer predicted leakers. More generally, if the SN that enriched the dwarf galaxies in the local Universe or the high-redshift galaxies have yields that are considerably different to what we have used in the simulations, our diagnostics will have a systematic error. Such an effect is expected to explain systematics shifts in the BPT diagram at $z\sim2-3$ \citep[e.g.][]{Steidel2016}.

Finally, it is important to note that there are some systematic differences between the simulated galaxies and those observed that impact the comparison. In particular, we have computed our {\small CLOUDY} models assuming isotropic radiation from the CMB.  To limit the computational demand, we have run the models assuming a CMB radiation field at $z=10$.  The CMB impacts the emergent luminosity of the [CII]~158$\mu$m line as well as the [OIII]~88$\mu$m line, particularly in lower density gas \citep[e.g.][]{dacunha2013,Lagache2018}.  Our model to identify LyC leakers based on IR lines takes a difference in the log luminosities of these lines so the impact of this is decreased to some extent and similarly, the bulk of the emission arises from the higher density gas that is impacted less; however, the DGS galaxies are not subject to this effect in the way that our simulated galaxies are, representing a systematic uncertainty.

\section{Conclusions}
\label{conclusion}
In this work, we have analysed high-redshift galaxies in the state-of-the-art Aspen simulations \citep{Katz2019a,Katz2019b} to determine whether lower-redshift LyC leakers \citep[e.g.][]{debarros2016,Rutkowski2017,Bian2017,Vanzella2016,Vanzella2018} are "analogues" of their epoch of reionization counterparts and to develop new emission line diagnostics for identifying LyC leakers. Our main conclusions are as follows:
\begin{itemize}
    \item Similar to observations, galaxies in our simulation that have high LyC $f_{\rm esc}$ are biased towards having high O32. We find that high O32 is likely a necessary but not sufficient condition for high $f_{\rm esc}$ in isolated galaxies (i.e. excluding mergers).
    \item Our models predict that there likely exist two populations in observations that have high O32 but low $f_{\rm esc}$.  The first are those where viewing angle causes a deviation of the line-of-sight $f_{\rm esc}$ from the angle averaged $f_{\rm esc}$, consistent with the explanation from \cite{Nakajima2019}. The second are those where a combination of metallicity and ionisation parameter allow for high O32 but low $f_{\rm esc}$, consistent with the models of \cite{Bassett2019}.  
    \item We find that the simulated high-redshift galaxies also populate the same regions of the R23-O32 plane as the $z\sim3$ LyC leakers presented in \cite{Nakajima2019} that tend to have metallicities $\lesssim0.5Z_{\odot}$. This, combined with evidence in the O32-$f_{\rm esc}$ plane suggests that these highly excited O32 emitters seem to be good analogues of galaxies in the epoch of reionization. 
    \item The LyC-emitting [SII]-deficient galaxies discussed in \cite{Wang2019} may not necessarily be analogues of high-redshift galaxies as strong differences in metallicity causes the simulated systems to populate a different region of the [SII] BPT diagram compared with the [SII]-deficient LyC leakers at low redshift. In general, the leakers in our simulations have [SII]-deficits but they also have deficits in [OIII] indicating that the dominant effect is metallicity/mass rather than a property of the ISM.
    \item We present multiple logistic regression models to better select LyC leakers from observational data. A minimal model that uses only the ${\rm [CII]_{158\mu m}}$ and ${\rm [OIII]_{5007\angstrom}}$ lines and has a true positive rate of 40\% while this value increases to 50\% when six lines are included in our most complex models.  We intend these models to be used to more efficiently identify LyC leaker candidates at low redshift from existing and future surveys.
    \item We present an IR-only model that uses the ${\rm [CII]_{158\mu m}}$ and ${\rm [OIII]_{88\mu m}}$ to classify galaxies that have $f_{\rm esc}\geq10\%$ and successfully apply this to local galaxies from the Dwarf Galaxy Survey \citep{Madden2013}, as well as various high-redshift galaxies that have been observed deep into the epoch of reionization.
    \item Based on our results from the IR-only model, MACS1149\_JD1 is the best galaxy candidate that has currently been observed that is likely to be actively contributing to the reionization of the Universe.
\end{itemize}

In addition to being a powerful probe of ISM properties, emission lines are a useful indicator for the escape of ionising photons in both low- and high-redshift galaxies. In future work, we will extend the models presented here to include Ly$\alpha$ which has been determined, both observationally and theoretically, to be a good indicator of $f_{\rm esc}$ as well as look how O32 correlates with $f_{\rm esc}$ on cloud scales not resolved by the current simulations.

\section*{Acknowledgements}
We thank the referee, Prof. Akio Inoue, for their comments which greatly improved the manuscript. This work made considerable use of the open source analysis software {\small PYNBODY} \citep{Pontzen2013}.  TK was supported in part by the National Research Foundation of Korea (NRF-2017R1A5A1070354 and NRF-2020R1C1C100707911) and in part by the Yonsei University Future-leading Research Initiative (RMS2-2019-22-0216). Support by ERC Advanced Grant 320596 ``The Emergence of Structure during the Epoch of reionization" is gratefully acknowledged by MH.  JB and JR acknowledge support from the ORAGE project from the Agence Nationale de la Recherche under grant ANR-14-CE33-0016-03. The research of AS and JD is supported by the Beecroft Trust and STFC.  RSE acknowledges funding from the European Research Council (ERC) under the European Union's Horizon 2020 research and innovation programme (grant agreement No 669253). NL acknowledges support from the Kavli foundation.

This work was performed using the DiRAC Data Intensive service at Leicester, operated by the University of Leicester IT Services, which forms part of the STFC DiRAC HPC Facility (www.dirac.ac.uk). The equipment was funded by BEIS capital funding via STFC capital grants ST/K000373/1 and ST/R002363/1 and STFC DiRAC Operations grant ST/R001014/1. DiRAC is part of the National e-Infrastructure.

\section*{Data Availability}
The data underlying this article will be shared on reasonable request to the corresponding author.

\bibliographystyle{mnras}
\bibliography{refs} 

\bsp	
\label{lastpage}
\end{document}